\documentclass{book}
\usepackage[contrib,lang=british]{ems-book} %% change to `american' if you use American English
\usepackage[sort,compress,numbers]{natbib}
\newcommand{\bit}{\begin{itemize}}
\newcommand{\eit}{\end{itemize}}
\def\bm#1{\mbox{\boldmath $ #1 $}}
\DeclareMathOperator{\tr}{tr}
\def\Rb{\overline {\mathbb R}}
\begin{document}
\mainmatter

%------
% Insert the title of your paper and (if necessary)
% a short title for the running head.
%------
\title{Lieb variation principle in density-functional theory}
\titlemark{Lieb variation principle in density-functional theory}

%------
% Insert full names of the authors.
% Add further authors as follows:
%  \emsauthor{2}{}{}
%  \emsauthor{3}{}{}
% etc.
% Abbreviate first names for the running head.
%------
\emsauthor{1}{Trygve Helgaker}{T.~Helgaker}
\emsauthor{2}{Andrew M. Teale}{A.~M.~Teale}
%------
% Use \authormark if the list of authors is too
% long for the running head: \authormark{A.~Doe et al.}
%------

%------
% Add one \emsaffil and one \email for each author.
% NOTE: The address does NOT appear in the paper.
% It will probably be printed in an appendix.
%------
\emsaffil{1}{Hylleraas Centre for Quantum Molecular Sciences, Department of Chemistry, University of Oslo, P.O.Box 1033, N-0315 Oslo, Norway \email{t.u.helgaker@kjemi.uio.no}}

%------
% Add MSC 2020 codes according to www.ams.org/msc/msc2020.html.
% Secondary codes (in square brackets) are optional.
%------
\classification[YYyYY]{XXxXX}

%------
% Add a list of keywords.
%------

%------
% Insert your abstract.
%------
\begin{abstract}
Lieb's convex formulation of density-functional theory is presented in a pedagogical manner, emphasizing its connection to Hohenberg--Kohn theory and to Levy's constrained-search theory. The Hohenberg--Kohn and Lieb variation principles are discussed, highlighting the dual relationship between the ground-state energy and the universal density functional.
%and relating the nondifferentiability of the density functional to the nonstrict concavity of the ground-state energy. 
Applications of the Lieb variation principle are reviewed, demonstrating how it may be utilized to calculate the Kohn--Sham potential of atoms and molecules, to study the exchange--correlation functional and the adiabatic connection by high-precision many-body methods, and to calculate the exchange--correlation hole and energy densities of atoms and molecules.

\end{abstract}

\makecontribtitle

%------
% INSERT THE BODY OF THE PAPER HERE (except
% acknowledgments, funding info and bibliography)
%------

Density-functional theory (DFT) is a hugely successful and powerful theory, underlying nearly all electronic-structure calculations of molecules and materials in chemistry and physics today. It is also a very beautiful theory -- an elegant application of convex analysis to electronic-structure theory, predicated on the simple observation that the ground-state electronic energy of an atom, molecule, or material is continuous and concave in the external potential, as realized in 1983 by Lieb in his seminal work `Density Functionals for Coulomb Systems'~\cite{Lieb1983}. Unfortunately, nearly forty years later, Lieb's convex formulation of DFT is still not widely appreciated among the practitioners of DFT or even among many of its developers. Likewise, the teaching of DFT is still mostly based on the Hohenberg--Kohn theorem \cite{HohKoh1964} and Levy's constrained-search theory \cite{Levy1979}. This is a pity since Lieb's convex, unifying formulation of DFT is both accessible and transparent -- and, in our experience, likely to intrigue students. 

Lieb's theory does not only give valuable insight into DFT. It also furnishes us with a practical tool for computation: the Lieb variation principle, according to which we may calculate to high accuracy, for any given electron density, the corresponding external potential and the universal density functional, thereby providing us with a practical realization of the Hohenberg--Kohn mapping from densities to potentials. Furthermore, it allows us to study the adiabatic connection and to calculate the universal density functional of DFT using high-precision quantum-chemical methods, providing both insight and valuable benchmark data for the development of approximate density functionals.

The bulk of this chapter consists of two parts. We begin in Section\,\ref{secA} by reviewing Lieb's convex formulation of DFT, emphasizing its relationship to Hohenberg--Kohn theory and to Levy's constrained-search theory. Next, in Section\,\ref{secB}, we review and illustrate the Lieb variation principle as a computational tool in DFT, with applications to  both Kohn--Sham theory and orbital-free DFT. Section~\ref{conclusions} contains some concluding remarks.

\section{Lieb's convex formulation of density-functional theory}
\label{secA}

%In this section, we present and discuss the convex formulation of DFT pioneered by Lieb in 1983~\cite{Lieb1983}. 

The following presentation of Lieb's formulation of DFT
is intended to be an informal introduction rather than a rigorous exposition of the theory. Concepts and elements of convex analysis are introduced as needed, in a hopefully self-contained manner. For an accessible, rigorous treatment convex analysis, we highly recommend the introductory text of van Tiel \cite{vanTiel1984}.

\subsection{Hohenberg--Kohn theorem}
\label{HKT1}

The Hamiltonian of an atomic or molecular system of $N$ electrons in an external potential $v$ can be written in the  form:
\begin{equation}
H(v) = T + W + \sum\nolimits_{i=1}^N v(\mathbf r_i). \label{eq:Ham1}
\end{equation}
Whereas the external potential $v$ varies from system to system, the kinetic operator $T = - \frac{\hbar^2}{2m_\text e} \sum_{i=1}^N \nabla_{{\mathbf r}_i}^2$ and the electron-repulsion operator $W=\frac{e^2}{4\pi \varepsilon_0}\sum_{i=1}^N \sum_{j=1}^i  \vert \mathbf r_i - \mathbf r_j \vert^{-1}$ are the same for all $N$-electron systems.  We are particularly interested in those potentials $v$ that support an electronic ground state and denote the set of these potentials by $\mathcal V_N$. 

For a given $v \in \mathcal V_N$, the ground-state energy $E(v)$ is obtained by solving the {Schr\"odinger equation},
\begin{equation}
H(v) \Psi(\mathbf x_i)  = E(v) \Psi (\mathbf x_i). \label{schrodinger}
\end{equation}
 Its solution is a complicated many-body problem,  the wave function $ \Psi(\mathbf x_i)$ depending on the spatial and spin coordinates $\mathbf x_i = (\mathbf r_i, \sigma_i)$ of all electrons.
By contrast, the associated ground-state density is a much simpler quantity, depending only on three spatial coordinates:
\begin{equation}
\rho (\mathbf r) = \int\! \vert \Psi(\mathbf r, \sigma, \mathbf x_2, \dots \mathbf x_N)\vert^2 \, \mathrm d \sigma \mathrm d \mathbf x_2 \cdots \mathrm d \mathbf x_N,
\end{equation}
where we interpret the integration over spin coordinates as a summation over $\alpha$ and $\beta$ spins.
According to the \emph{Hohenberg--Kohn theorem}, we may use the density as a fundamental variable in place of the wave function when studying atomic and molecular systems \cite{HohKoh1964}:
\begin{quote}`Thus $v(\mathbf r)$ is (to within a constant) a unique functional of $\rho(\mathbf r)$; since, in turn,
$v(\mathbf r)$ fixes $H(v)$ we see that the full many-body ground state is a unique functional of $\rho(\mathbf r)$.'
\end{quote}
Equipped with this theorem, we can  set up the \emph{Hohenberg--Kohn mapping} from ground-state densities to ground-state wave functions via the external potential and the Hamiltonian:
\begin{equation}
\rho \mapsto v_\rho \mapsto H(v_\rho) \mapsto \Psi_\rho.
\end{equation}
A density that is the ground-state density for some potential $v \in \mathcal V_N$ is said to be \emph{$v$-representable} and the set of all $v$-representable densities is denoted by $\mathcal A_N$.

The Hohenberg--Kohn theorem (1964) has played an enormously important role in theoretical chemistry -- indeed, it is typically viewed as the cornerstone of DFT, taught in all introductions to DFT along with Levy's constrained-search theory (1979). At the same time, the Hohenberg--Kohn theorem may appear almost mysterious and difficult to understand intuitively. Although Levy's constrained-search formulation brings  clarity, the underlying structure and beauty of DFT is best appreciated from the point of view Lieb's convex formulation of the theory (1983).
%\cite{Lieb1983}.

It is interesting to note that many of the elements or concepts of Lieb's theory were already present in the early work of Hohenberg and Kohn but were not picked up on and developed further at the time. We will here highlight these connections, beginning with the subgradient inequality of the ground-state energy.

\subsection{Subgradient inequality of the ground-state energy}
\label{sec112}

\begin{figure}
\centering
{\scalebox{0.7}{\includegraphics{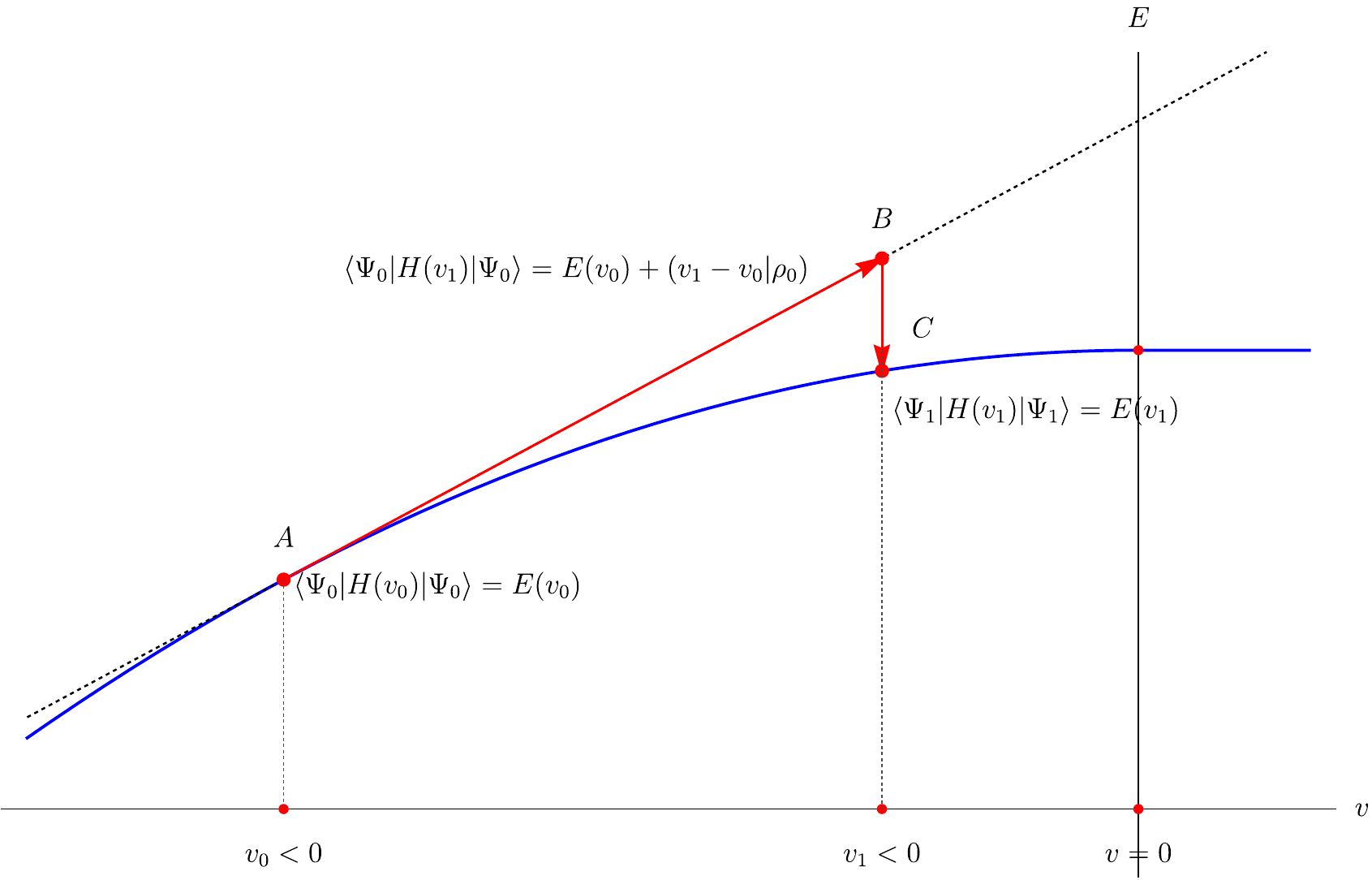}}}
\caption{The concavity of the ground-state energy $v \mapsto E(v)$ in the potential $v$ (blue line),
assuming that the potentials $v_0,v_1 \in \mathcal V_N$ have ground-state wave functions
$\Psi_0$ and $\Psi_1$, respectively. As the potential changes from $v_0$ to $v_1$ with the ground-state wave function $\Psi_0$ fixed along the red line, the energy
changes linearly. When the wave function relaxes from $\Psi_0$ to the ground-state $\Psi_1$ at $v_1$, the ground-state
energy is lowered, generating a concave curve $v \mapsto E(v)$. The slope of the tangent to $E$ at $v_0$ is the density of the ground-state wave function $\Psi_0$ at $v_0$. In the terminology of convex analysis, the ground-state density $\rho_0$ is the `subgradient' of the ground-state at the potential $v_0$. The `subgradient inequality' in Eq.\,\eqref{ineq1} therefore expresses the notion that the straight red line in the plot is a tangent to the concave blue curve.}
\label{figEconcave}
\end{figure}

As part of the proof of the Hohenberg--Kohn theorem\,\cite{HohKoh1964}, the following \emph{subgradient inequality} of the ground-state energy at $v \in \mathcal V_N$ was established:
\begin{equation}
\begin{alignedat}{2}
E(u) &< E(v) + ( u - v \,\vert\, \rho_v), \quad& \forall u &\notin v + \mathbb R,
\\
E(u) &= E(v) + ( u - v \,\vert\, \rho_v), \quad &\forall u &\in v + \mathbb R.
\end{alignedat}
\label{ineq1}
\end{equation}
Here $\rho_v \in \mathcal A_N$ is a ground-state density associated with $H(v)$ and we use the short-hand notation
\begin{equation}
(u \vert \rho) = \int\!\!\! u(\mathbf r) \rho(\mathbf r)\, \mathrm d \mathbf r
\end{equation}
to denote the interaction of $u$ with $\rho$. 
An intuitive understanding of the subgradient inequality can be obtained from Figure\,\ref{figEconcave}, which shows how the inequality arises from the Rayleigh--Ritz variation principle and the linearity of the Hamiltonian in the potential. As seen in the figure, the subgradient inequality is a consequence of the \emph{concavity of the ground-state energy}. Geometrically speaking, a function is said to be concave if every chord connecting two points on its graph lies on or below the graph. As we shall see later, concavity of the ground-state energy is the key to DFT.
 
%We shall later see that the subgradient inequality is a consequence of the concavity of the ground-state energy.

The subgradient inequality gives the Hohenberg--Kohn theorem directly. Let the potentials $u,v \in \mathcal V_N$  differ by more than an additive constant and let
$\rho_u,\rho_v \in \mathcal A_N$ be the corresponding ground-state densities.
From Eq.\,\eqref{ineq1}, we then obtain the strict subgradient inequalities $E(u) < E(v) + (u-v \,\vert\, \rho_v)$ and $E(v) < E(u) + (v-u \,\vert\, \rho_u)$. Adding these together, we conclude that $(v - u \,\vert \, \rho_v - \rho_u) < 0$ and hence that $\rho_u \neq \rho_v$ in accordance with the Hohenberg--Kohn theorem. The Hohenberg--Kohn theorem is thus a simple consequence of the (strict) concavity of the ground-state energy in the external potential, which in turn follows from the Rayleigh--Ritz variation principle and the linearity of the Hamiltonian in the potential.

\subsection{Hohenberg--Kohn and Lieb variation principles}
\label{HKT2}

The subgradient inequality of the ground-state energy is also the key to the Hohenberg--Kohn variation principle.
For a $v$-representable density $ \rho \mapsto v_\rho$, the \emph{Hohenberg--Kohn universal density functional} \cite{HohKoh1964} is defined as
\begin{equation}
F_\text{HK}(\rho) = E(v_\rho) - (v_\rho \vert \rho),
\end{equation}
where $F_\text{HK}\colon \mathcal A_N \to \mathbb R$ and
$E\colon \mathcal V_N \to \mathbb R$. Combining this definition of $F_\text{HK}$ with the subgradient inequality of Eq.\,\eqref{ineq1} expressed as $E(v_\rho) - ( v_\rho \vert \rho) \geq E(v)  - (v \vert \rho)$, we obtain
\begin{equation}
F_\text{HK}(\rho) \geq  E(v) - (v \vert \rho), \label{Fhk}
\end{equation}
valid for every $\rho \in \mathcal A_N$ and for every $v \in \mathcal V_N$. 
%For each $v \in \mathcal V_N$, the
From this inequality, the \emph{Hohenberg--Kohn and Lieb variation principles}, respectively, follow directly:
\begin{alignat}{2}
E(v) &= \min_{\rho \in \mathcal A_N} \left (F_\text{HK}(\rho) + (v \vert \rho) \right), &\quad \forall v &\in \mathcal V_N, \label{HKvp0} \\
F_\text{HK}(\rho) &= \max_{v\in \mathcal V_N} \left(E(v) - (v \vert \rho)\right), &\quad \forall \rho &\in \mathcal A_N. \label{Lvp0}
\end{alignat}
We may thus calculate the ground-state energy from the universal density functional by the Hohenberg--Kohn variation principle~\cite{HohKoh1964}; conversely, the universal density functional is obtained from the ground-state energy by the Lieb variation principle. Together, these variation principles highlight a symmetry or duality between potentials and ground-state densities and also between the ground-state energy and the universal density functional. However, this duality was not emphasized in the work by Hohenberg and Kohn, who  considered only the first variation principle~Eq.\,\eqref{HKvp0}. Lieb was the first to study both variation principles rigorously, within the framework of convex analysis~\cite{Lieb1983}.

We note the \emph{$v$-representability problem} of Hohenberg--Kohn theory and the variation principles established above: the sets $\mathcal A_N$ and $\mathcal V_N$ do not form vector spaces and are not explicitly known.
Also,
we have no optimality conditions except by the Hohenberg--Kohn mapping $\rho \mapsto v_\rho$. These restrictions will be lifted with the development of the constrained-search formalism of Levy~\cite{Levy1979} and the convex formulation by Lieb~\cite{Lieb1983}.
%\eit

%\eit
%\item We have so far \alert{no optimality conditions} except by the HK mapping $\clb \rho \mapsto v_\rho$
%\bit \scriptsize
%It is our purpose to explore the optimality conditions in more detail.
%\subsection{Optimality conditions}

%At optimality, the Hohenberg--Kohn and Lieb variation principles satisfy the conditions
%$E(v) = F_\text{HK}(\rho) + (v \vert \rho)$ and $F_\text{HK}(\rho) = E(v) - (v \vert \rho)$ 
%from which it follows that
%\begin{alignat}{2}F_\text{HK}(\rho) + (v \vert \rho) &\leq F_\text{HK}(\tilde \rho) + (v \vert \tilde \rho) , &\quad \forall \tilde \rho &\in \mathcal A_N, \label{HKvp01} \\E(v) - (v \vert \rho) &\geq E(\tilde v) - (\tilde v \vert \rho) , &\quad \forall \tilde v &\in \mathcal V_N. \label{Lvp01}
%\end{alignat}
%which may be rewritten in the following  form
%\begin{alignat}{2} F_\text{HK}(\tilde \rho) &\geq F_\text{HK}(\rho) + (-v \vert \tilde \rho - \rho)  , &\quad \forall \tilde \rho &\in \mathcal A_N, \label{HKvp02} \\E(\tilde v) &\leq E(v) + (\tilde v - v \vert \rho) , &\quad \forall \tilde v &\in \mathcal V_N. \label{Lvp02}
%\end{alignat}
%A potential $-v$ that satisfies Eq.\,\eqref{HKvp02} is said to be a subgradient of $F_\text{HK}$ at $\rho$,while a density $\rho$ that satisfies Eq.\,\eqref{Lvp02} is said to be a supergradient of $E$ at $v$.

\subsection{Vector spaces of densities and potentials}

%Hohenberg--Kohn theory works with the set of ground-state densities $\mathcal A_N$  and the set of potentials that support ground-state densities $\mathcal V_N$. 
To proceed, we will need to work with vector spaces that contain all densities and all potentials of interest to us. Lieb introduced
the complete vector spaces
\begin{equation}
X = L^3(\mathbb R^3) \cap L^1(\mathbb R^3), \quad
X^\ast= L^{3/2}(\mathbb R^3) + L^\infty(\mathbb R^3) \label{X}
\end{equation}
where the space of potentials $X^\ast$ is dual to the space of densities $X$, thereby ensuring that all interactions $(v \vert \rho)$ with $\rho \in X$ and $v \in X^\ast$ are finite. While $X^\ast$ is sufficiently large to include all Coulomb potentials, $X$ contains all normalized densities arising from $N$-electron states of a finite kinetic energy -- such densities are said to be \emph{$N$-representable}. The set of $N$-representable densities $\mathcal I_N$ can 
be characterized in a simple manner,
\begin{equation}
\mathcal I_N = \{ \rho \in X \mid \rho \geq 0, \, \textstyle \int\!\!\rho(\mathbf r) \,\mathrm d \mathbf r = N, \, T_\text{W}(\rho) < + \infty \}, 
\end{equation}
where $T_\text W \colon \mathcal I_N \to [0,+\infty]$ is the \emph{von Weizs\"acker kinetic energy functional}:
\begin{equation}
T_\text{W}(\rho) =
\tfrac{1}{2} \!\int\! \vert \bm \nabla \rho^{1/2}(\mathbf r) \vert^2 \, \mathrm d \mathbf r .
\label{TW}
\end{equation}
 It can be shown that $\mathcal I_N$ is a \emph{convex set} and that $T_\text W$ is a \emph{convex function}, meaning that, for each pair of different densities $\rho_1,\rho_2 \in \mathcal I_N$ and every $\lambda \in (0,1)$, the \emph{convex combination} $\lambda \rho_1 + (1-\lambda) \rho_2$ also belongs to $\mathcal I_N$ and that $T_\text W$ satisfies the \emph{convexity characterization} of a convex function:
\begin{equation}
T_\text W(\lambda \rho_1 + (1-\lambda) \rho_2 ) \leq \lambda T_\text W(\rho_1) + (1-\lambda) T_\text W(\rho_2) .
\end{equation}
We note that $\mathcal A_N \subsetneq \mathcal I_N \subsetneq X$.
%that $E(v)$ is well-defined and finite for each $v \in X^\ast$,
%and (3) that $X^\ast$ contains all Coulomb potentials. 

Even though $X^\ast$ contains all Coulomb potentials, it does not contain all potentials that may be of interest to us -- harmonic potentials, for example, support an electronic ground state but are not contained in $X^\ast$. Indeed, this limitation of the theory was already indicated in the title of Lieb's 1983 paper~\cite{Lieb1983}: `Density Functionals for Coulomb Systems'. The restriction is 
fairly mild, however, and does not reduce the usefulness of the theory significantly.
In keeping with this limitation of the theory, we redefine $\mathcal V_N$ to be the set of potentials in $X^\ast$ that support a ground state: $\mathcal V_N \subsetneq X^\ast$.

\subsection{Levy--Lieb constrained-search functional}
\label{secLL}

While Hohenberg--Kohn theory is based on the Schr\"odinger equation, the constrained-search theory is based on the 
\emph{Rayleigh--Ritz variation principle} for the ground-state energy $E\colon X^\ast \to \mathbb R$, which for pure states may be written as:
\begin{equation}
E(v) = \inf_{\Psi \mapsto N} \langle \Psi \vert H(v) \vert \Psi \rangle.
\label{RR}
\end{equation}
The ground-state energy $E(v)$ is well defined (finite) for each $v \in X^\ast$ provided the minimization is over all antisymmetric $N$-electron wave functions of a finite kinetic energy, as indicated by the notation $\Psi \mapsto N$. We note, however, that the infimum is not always achieved -- for the oxygen atom, for example, a minimum is obtained for atoms containing up to nine electrons, but no minimum exists for more than nine electrons. In other words, the potential $v_Z(\mathbf r) = - Ze^2/(4 \pi \varepsilon_0 r)$ with $Z=8$ belongs to $\mathcal V_N$ when $N \leq 9$ but not when $N > 9$.
Nevertheless, $E(v)$ is given by Eq.\,\eqref{RR} is well defined in all cases and we use the term `ground-state energy' even when the infimum is not achieved and no ground state exists.

Following the seminal work of Levy~\cite{Levy1979}, we now rewrite the Rayleigh--Ritz variation principle of Eq.\,\eqref{RR} in the manner
\begin{equation}
E(v) = \inf_{\rho \in \mathcal I_N} \inf_{\Psi \mapsto \rho} \langle \Psi \vert H(v) \vert \Psi \rangle,
\label{RR0}
\end{equation}
where the short-hand notation $\Psi \mapsto \rho$ indicates that the minimization is restricted to wave functions with density $\rho$.
Introducing the
\emph{Levy--Lieb constrained-search  functional} $F_\text{LL} \colon \mathcal I_N \to [0,+\infty)$ on the set of $N$-representable densities by
\begin{equation}
F_\text{LL}(\rho) = \inf_{\Psi \mapsto \rho} \langle \Psi \vert H(0) \vert \Psi \rangle, 
%\quad \rho \in \mathcal I_N,
\end{equation}
where $H(0) = T + W$,
we arrive at the following \emph{Hohenberg--Kohn variation principle} for $v \in X^\ast$~\cite{Levy1979}:
\begin{equation}
E(v) = \inf_{\rho \in \mathcal I_N} \left(F_\text{LL}(\rho) + (v \vert \rho) \right). \label{HKrr}
\end{equation}
Since $\mathcal I_N$ and $X^\ast$ are explicitly known, the $v$-representability problem of Hohenberg--Kohn theory has been solved. 

\subsection{Discontinuity of the universal density functional}

Although the constrained-search theory solves the $v$-representability problem, the
question regarding the optimality conditions in the Hohenberg--Kohn variation principle remains. Since $F_\text{LL}$ is defined on the vector space $X$,  
it is tempting to assume that we can use the Euler equations for this purpose, since we no longer need to worry about whether or not the functions in the immediate neighbourhood of optimal functions are $v$-representable:
\begin{equation}
\frac{\delta F_\text{LL}(\rho)}{\delta \rho(\mathbf r)} = - v(\mathbf r) + c, \quad c \in \mathbb R. \label{Euler}
\end{equation}
However, use of the Euler equations presumes that the density functional is differentiable. Unfortunately, differentiablity is precluded by the fact that the universal density-functional is  discontinuous~\cite{Lammert2007}.
%Since the constrained-search functionals are defined on complete vector spaces,it was initially assumed that one could use the Euler equations to identify the minimizers in the Hohenberg--Kohn variation principle by the conditions (and hence provide the Hohenberg--Kohn mapping from densities potentials):
%Regarding the optimality conditions of the Hohenberg--Kohn variation principle with the Levy--Lieb density functional, Levy writes:
%\begin{quote} `One can now confidently use existing Euler equations without being concerned about whether or not the functions
%in the immediate neighbourhood of the optimum functions are $v$-representable.'
%\end{quote}
%The Euler equations referred to by Lieb are given by
%\begin{equation}\frac{\delta F_\text{DM}(\rho)}{\delta\rho(\mathbf r)}  = -v(\mathbf r) + c, \quad c \in \mathbb R.
%\end{equation}
%and likewise for $F_\text{LL}$. It turns out, however, that the universal density functional is not differentiable and therefore that the Euler equations are not well defined. 

\begin{figure}
\centering
{\scalebox{0.85}{\includegraphics{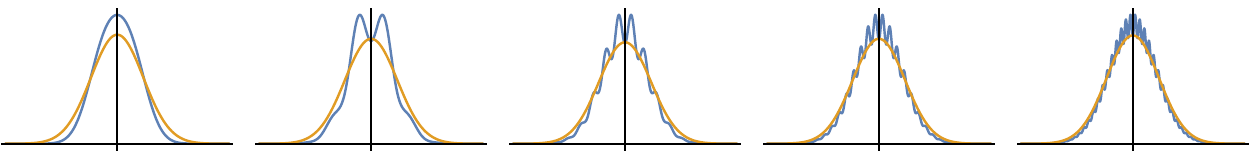}}}
\caption{A convergent sequence of densities %$\{\rho_n\}_{n=1}^\infty$ 
with a divergent von Weizs\"acker kinetic energy.}
%limit $\lim_{n\to \infty} \rho_n = \rho$ while $T_\text W(\rho_n)$ diverges.}
%while $\lim_{n\to \infty} T_\text W(\rho_n) = +\infty$.}
\label{TWfig}
\end{figure}

To illustrate, consider a one-electron system, for which the universal density functional has an explicit form, being equal to the von Weizs\"acker functional of Eq.\,\eqref{TW}.
A one-electron Gaussian density
of unit exponent has a finite kinetic energy:
\begin{equation}
\rho( \mathbf r) = \pi^{-3/2} \exp(-r^2), \quad T_\text{W}(\rho) = 3/4E_\text h.
\end{equation}
Let now $\{\rho_n\}_{n=1}^\infty$ be a sequence that approaches $\rho$ in the norm of $X$:
\begin{equation}
\lim_{n\to +\infty} \Vert \rho - \rho_n \Vert_X = 0,
\end{equation}
while developing increasingly rapid oscillations of increasingly small amplitude, as illustrated in Figure\,\ref{TWfig}.
The kinetic energy $T_\text{W}(\rho_n)$ is then driven arbitrarily high in the sequence, implying that $T_\text W$ is not continuous:
\begin{equation}
\quad \lim_{n \to \infty} T_\text{W}\left(\rho_n\right) = + \infty   \neq  T_\text{W}\left(\lim_{n \to \infty}\rho_n\right) = 3/4E_\text h.
\end{equation}
Since it can be shown that $T_\text W \leq F_\text{LL}$ for each $N \geq 1$,
this result also means that $F_\text{LL}$ is  discontinuous and hence not differentiable. Consequently, the Euler equations in Eq.\,\eqref{Euler} are not well defined. We will return to the problem of identifying minimizers in the Hohenberg--Kohn variation principle later, when the proper tools have been developed.
%(P. E. Lammert, {\it Int.\ J. Quantum Chem.}\ {\bf 107}, 1943 (2007)).

%It can be shown that a minimizing densities is always found in the Levy--Lieb constrained-search functional, which can thereforebe written as
%\begin{equation}
%F_\text{LL}(\rho) = \min_{\Psi \mapsto \rho} \langle \Psi \vert T + W \vert \Psi \rangle.
%\end{equation}
%This result has important consequences, as we shall discuss later. 

%In particular, it implies that the infimum is achieved for a given $v \in X^\ast$ in the Hohnenberg--Kohn variation principle if and only if the infimum is achieved in the Rayleigh--Ritz variation principle and that the minimizing densities in the Hohenberg--Kohn variation principle are precisely the (pure-state) ground-state densities in the Rayleigh--Ritz variation principles. DFT is therefore faithful: no ground-state densities are lost nor does any spurious ground-state densities occur.

\subsection{Rayleigh--Ritz variation principle for ensembles}

%In Section\,\ref{secLL}, we discussed constrained-search theory functional in terms of pure state, introducing the Levy--Lieb constrained-search functional $F_\text{LL}$. In many ways, a
%more convenient formulation of constrained-search theory is in terms of ensembles.
Our next step is to generalize the constrained-search theory to ensembles, taking as our starting point the
\emph{canonical-ensemble Rayleigh--Ritz variation principle}:
\begin{equation}
E(v) = \inf_{\gamma \to N} \tr \left(\gamma H(v)\right).
\label{enRR}
\end{equation}
Here each density matrix $\gamma$ is a normalized self-adjoint trace-class operator on the $N$-electron Hilbert space, which
may be given the spectral decomposition
\begin{equation}
\gamma = \sum_{i=1}^\infty \lambda_i \vert \Psi_i \rangle \langle \Psi_i \vert, \quad \lambda_i \geq 0, \quad \sum_{i=1}^\infty \lambda_i = 1, \end{equation}
in terms of an orthonormal basis $\{\Psi_i\}_{i=1}^\infty$. The expectation value in Eq.\,\eqref{enRR} may then be written as
\begin{equation}
\tr \left(\gamma H(v) \right) = \sum_{i=1}^\infty \lambda_i  \langle \Psi_i \vert H(v) \vert \Psi_i \rangle.
\end{equation}
Each $M$-degenerate ensemble ground-state density matrix obtained from Eq.\,\eqref{enRR} using $v \in \mathcal V_N$ is a (finite) convex combination of $M$ pure ground-state density matrices,
\begin{equation}
\gamma_0 = \sum_{i=1}^M \lambda_{0i} \vert \Psi_{0i} \rangle \langle \Psi_{0i} \vert, \quad \lambda_{0i} \geq 0, \quad \sum_{i=1}^M \lambda_{0i} = 1,
\end{equation}
where each $\Psi_{0i}$ is a ground-state wave function obtained from Eq.\,\eqref{RR} with the same potential $v$. By the same token, each ensemble ground-state density is a convex combination of pure ground-state densities with the same potential. 
Generalizing the concept of $v$-representability to ensembles, we say that
a density $\rho \in \mathcal I_N$ is \emph{ensemble $v$-representable} if it is the ensemble ground-state density for some external potential $v \in \mathcal V_N$. The set of all ensemble $v$-representable densities is denoted by $\mathcal B_N$. Clearly, $\mathcal A_N \subsetneq \mathcal B_N$.

It should be emphasized that the Rayleigh--Ritz variation principle for pure states and the
Rayleigh--Ritz variation principle for ensembles give the same ground-state energy for every $v \in X^\ast$ -- the only difference is that more solutions (minimizing density matrices) are obtained with the ensemble variation principle.

\subsection{Lieb constrained-search functional}

Proceeding as for pure states, we obtain from the ensemble Rayleigh--Ritz variation principle in Eq.\,\eqref{enRR} the \emph{ensemble Hohenberg--Kohn variation principle}
\begin{equation}
E(v) = \inf_{\rho \in \mathcal I_N} \left(F_\text{DM}(\rho) + (v \vert \rho) \right),
\label{EFdm}
\end{equation}
where the \emph{Lieb density-matrix constrained-search functional} \cite{Lieb1983} is given by
\begin{equation}
F_\text{DM}(\rho) = \inf_{\gamma \mapsto \rho} \tr( \gamma H(0)).
\label{Fdminf}
\end{equation}
A nontrivial result due to Lieb~\cite{Lieb1983} is that the infimum is always achieved,
\begin{equation}
F_\text{DM}(\rho) = \min_{\gamma \mapsto \rho} \tr( \gamma  H(0))
= \tr( \gamma_\rho  H(0)),
\label{Fdm}
\end{equation}
with important implications for the theory. A similar result holds for $F_\text{LL}$.
%where we have taken the opportunity to replace infimum by minimum, noting that a minimizing density matrix exists for each $\rho \in \mathcal I_N$.

Since the constrained search over ensembles in $F_\text{DM}$ includes the constrained search over pure states in $F_\text{LL}$, we conclude that
\begin{equation}
F_\text{DM} \leq F_\text{LL}.
\end{equation}
In fact, $F_\text{DM} \neq F_\text{LL}$, since there exist densities $\rho \in \mathcal I_N$ for which $F_\text{DM}(\rho) < F_\text{LL}(\rho)$.
Nevertheless, both functionals are \emph{admissible}, meaning that both
give the correct ground-state energy for every potential in the Hohenberg--Kohn variation principle. They differ only in that $F_\text{DM}$ gives more minimizing solutions -- specifically, each minimizing density obtained with $F_\text{DM}$ is a convex combination of those obtained with $F_\text{LL}$. The reason for this behaviour is that $F_\text{DM}$ is the \emph{convex hull} of $F_\text{LL}$ -- that is, its greatest convex lower bound \cite{Lieb1983}.

\text{Convexity} is a powerful property of $F_\text{DM}$, with far-reaching consequences for the theory. The proof is simple. Let $\rho_1,\rho_2 \in \mathcal I_N$ be different and let $\lambda \in (0,1)$.
%, the following inequality is satisfied:
%\begin{equation}
%F_\text{DM}(\lambda \rho_1 + (1- \lambda) \rho_2) \leq \lambda F_\text{DM}(\rho_1) + (1-\lambda) F_\text{DM}(\rho_2). \label{Fconvex}
%\end{equation}
%The proof of this \emph{interpolation characterization} is simple. 
There then exist by Eq.\,\eqref{Fdm} minimizing density matrices
 $\gamma_1 \mapsto \rho_1$ and $\gamma_2 \mapsto \rho_2$  in $F_\text{DM}$ such that 
$F_\text{DM}(\rho_1) = \tr( \gamma_1 H(0))$ and $F_\text{DM}(\rho_2) = \tr( \gamma_2 H(0))$. 
By the definition of $F_\text{DM}$, we then have
\begin{align}
F_\text{DM}(\lambda \rho_1 + (1-\lambda)\rho_2) &=
 \min\nolimits_{\gamma \to \lambda \rho_1 + (1-\lambda)\rho_2} \tr( \gamma H(0)) \nonumber \\
&\leq \tr( (\lambda \gamma_1 + (1-\lambda) \gamma_2 )H(0)), \label{Fdm31}
%\nonumber \\ &= \lambda F_\text{DM}(\rho_1) + (1-\lambda) F_\text{DM}(\rho_2)
\end{align}
where the inequality holds since $\lambda \gamma_1 + (1-\lambda) \gamma_2 \mapsto \lambda \rho_1 + (1-\lambda) \rho_2$ is an allowed (but not necessarily minimizing) density matrix in the constrained search. Reintroducing $F_\text{DM}$ on the right-hand side of Eq.\,\eqref{Fdm31}, we obtain
\begin{equation}
F_\text{DM}(\lambda \rho_1 + (1- \lambda) \rho_2) \leq \lambda F_\text{DM}(\rho_1) + (1-\lambda) F_\text{DM}(\rho_2). \label{Fconvex}
\end{equation}
verifying convexity.

Another important property of $F_\text{DM}$ is \emph{lower semi-continuity} \cite{Lieb1983}, meaning that, for each $\rho \in X$ and each sequence $\{\rho_n\}_{n=1}^{\infty}$ converging to $\rho$, we have
\begin{equation}
\liminf_{n \to\infty} F_\text{DM}(\rho_n) \geq F_\text{DM}(\rho).
\end{equation}
Roughly speaking, lower semi-continuity at $\rho$ means that $F_\text{DM}$ may jump up but not down as we move away from $\rho$. 
Functions that are both convex and lower semi-continuous are said to be \emph{closed convex}, an important, well-behaved class of functions to which $F_\text{DM}$ belongs.

\subsection{Minimizing densities}
%in the Hohenberg--Kohn variation principle}

A useful property of convex functions is that every local minimum is a global minimum, greatly simplifying their minimization. Since $F_\text{DM}$ is convex, $\rho \mapsto F_\text{DM} + (v \vert \rho)$ is also convex, implying that \emph{every local minimizing density in the (ensemble) Hohenberg--Kohn variation principle is a global minimizer}. Moreover, for a given potential $v \in X^\ast$, the set of minimizers constitute a convex set, which is nonempty if and only if $v \in \mathcal V_N$. 
%A set $A$ is a \emph{convex set} if, for each $a_1,a_2 \in A$ and each $\lambda \in (0,1)$, the convex combination $\lambda a_1 + (1-\lambda) a_2$ also belongs to $A$.

Let us consider the relationship of the minimizing densities in the Hohenberg--Kohn variation principle to the ground-state densities in
the Rayleigh--Ritz variation principle. Since $F_\text{DM}(\rho) + (v \vert \rho) = \min_{\gamma \mapsto \rho} \tr (\gamma H(v))$, we may write
\begin{equation}
E(v) = \begin{cases}\inf_{\gamma }  \tr ( \gamma H(v)), &  \mbox{Rayleigh--Ritz variation principle,} \\
\inf_{\rho \in \mathcal I_N} \min_{\gamma \to \rho}  \tr ( \gamma H(v)), &  \mbox{Hohenberg--Kohn variation principle,}
\end{cases}
\label{1.25}
\end{equation}
%Since a minimizing density matrix exists for each $\rho \in \mathcal I_N$ in $F_\text{DM}$, we may rewrite the objective function in the Hohenberg--Kohn variation principle in the manner
%\begin{equation}\rho \mapsto F_\text{DM}(\rho) + (v \vert \rho) 
%= \min_{\gamma \mapsto \rho} \tr (\gamma (T+W)) + (v\vert \rho)) 
%= \min_{\gamma \mapsto \rho} \tr (\gamma H(v)).
%\label{EFdminX}
%\end{equation}
%Using this expression, we may write the ground-state energy in the form
%\begin{equation}E(v) = \inf_{\rho \in \mathcal I_N} \min_{\gamma \to \rho}  \tr ( \gamma H(v)) = \inf_{\gamma }  \tr ( \gamma H(v)) 
%\label{1.25}
%\end{equation}
%where the first expression corresponds to the ensemble Rayleigh--Ritz variation principle and the second to the ensemble Hohenberg--Kohn variation principle. 
where the searches in the two variation principles are over the same complete set of density matrices. Therefore, if $v \in \mathcal V_N$,
then the infimum is achieved by the same density matrices in both cases:
\begin{equation}
E(v) = \min_{\gamma }  \tr ( \gamma H(v))
= \min_{\rho \in \mathcal I_N} \min_{\gamma \to \rho}  \tr ( \gamma H(v)), \quad v \in \mathcal V_N.
\label{mineq}
\end{equation}
It follows that \emph{the minimizing densities in the Hohenberg--Kohn variation principle are precisely the ground-state densities in the Rayleigh--Ritz variation principle}. We note how this result depends critically on the existence of a minimizing density matrix $\gamma \mapsto \rho$ in the constrained-search functional in Eq.\,\eqref{Fdm} and hence in Eq.\,\eqref{1.25}, precluding the existence of minimizing densities in the Hohenberg--Kohn variation principle that are not ground-state densities.
%are not ground-state densities.

In conclusion, the ensemble Hohenberg--Kohn variation principle is faithful in the sense that it gives not only the same ground-state energy as does the ensemble Rayleigh--Ritz variation principle, but also the same ground-state densities, when these exist. A similar result holds for pure states, with the Levy--Lieb functional.
%A similar but weaker result holds for $F_\text{LL}$. For given potential $v$, the global minimizers are precisely the pure ground-state densities of $H(v)$. However, since since $F_\text{LL}$ is not convex, a minimizing density may not be a global minimizer in the Hohenberg--Kohn variation principle with $F_\text{LL}$.

\subsection{The Lieb functional and variation principle}

Let us now develop DFT within the framework of convex analysis. 
%We have already noted that the ground-state energy $E\colon X^\ast \to \mathbb R$ in %Eq.\,\eqref{RR} is everywhere finite. 
The key to this development is the \emph{concavity} of the ground-state energy $E\colon X^\ast \to \mathbb R$ (alluded to in Section\,\ref{sec112}), meaning that the interpolation characterization
\begin{equation}
E(\lambda v_1 + (1- \lambda) v_2) \geq \lambda E(v_1) + (1-\lambda) E(v_2) \label{Econcave}
\end{equation}
holds for each pair $v_1,v_2 \in X^\ast$ and each $\lambda \in (0,1)$. 
%We note that a function is concave if and only if minus the function is convex. %Concavity implies that a linear interpolation of the ground-state energy between two potentials never overestimates the true energy.
%Concavity of the ground-state energy is the key to DFT and it is worthwhile to demonstrate it explicitly.
To demonstrate concavity, we first note that $H(\lambda v_1 + (1-\lambda) v_2) = \lambda H(v_1) + (1-\lambda) H(v_2)$ holds by the linearity of $v \mapsto H(v)$.
We may therefore rewrite the Rayleigh--Ritz variation principle (here in a pure-state formulation) as:
%\begin{align} E(\lambda v_1 + (1-\lambda) v_2)  &= \inf_\gamma \left(\tr \gamma H(\lambda v_1 + (1-\lambda) v_2) \right) \nonumber \\
%&= \inf_\gamma \left(\lambda \tr \left(\gamma H(v_1) \right) + (1-\lambda) \tr \left(\gamma H( v_2) \right) \right) \label{Ecave1}\end{align}
\begin{align}
E(\lambda v_1 + (1-\lambda) v_2) 
&= \inf_\Psi \langle \, \Psi \,\vert\, H(\lambda v_1 + (1-\lambda) v_2) \,\vert\, \Psi \, \rangle \nonumber \\
&= \inf_\Psi \left(\lambda \langle \, \Psi \,\vert\, H(v_1) \,\vert\, \Psi \, \rangle + 
(1-\lambda)\langle \, \Psi \,\vert\, H(v_2) \,\vert\, \Psi \, \rangle \right).
\label{Ecave1}
\end{align}
Minimizing the two expectation values separately, we obtain
\begin{align}
E(\lambda v_1 + (1-\lambda) v_2) 
&\geq  \inf_\Psi \lambda \langle \, \Psi \,\vert\, H(v_1) \,\vert\, \Psi \, \rangle + \inf_\Psi(1-\lambda)\langle \, \Psi \,\vert\, H(v_2) \,\vert\, \Psi \, \rangle \nonumber \\
 &= \lambda \inf_\Psi \langle \, \Psi \,\vert\, H(v_1) \,\vert\, \Psi \, \rangle + 
 (1-\lambda)\inf_\Psi\langle \, \Psi \,\vert\, H(v_2) \,\vert\, \Psi \, \rangle ,
 \label{Ecave2}
\end{align}
where the last step follows since $\lambda$ and $1-\lambda$ are both nonnegative. Concavity of the ground-state energy is thus an immediate consequence of the Rayleigh--Ritz variation principle and of the linearity of the Hamiltonian in the external potential. We note that the proof of concavity does not carry through for excited states (except for the lowest state of each symmetry) since
the Rayleigh--Ritz minimization is then subject to orthogonality constraints that depend on the external potential, precluding the inequality in Eq.\,\eqref{Ecave2} from being established. 
%On the other hand, $v \mapsto \sum_{i=1}^n E_i(v)$ is concave if the summation is over all states of energy $E_i(v)$ less than or equal to $E_n(v)$, suggesting that a DFT-like theory for excited states can be set up based on such a sum over electronic states.

%Concavity does not hold for excited states (except for the lowest excited state of each symmetry) since the variational space for the wave function or the density matrix is then not the same for all external potentials in the Rayleigh--Ritz variation principle.

Since the ground-state energy is concave and everywhere finite, it is also continuous and hence \emph{closed concave}.
We may then apply the \emph{Fenchel--Moreau biconjugation theorem} of convex analysis to deduce the existence of a closed convex function $F\colon X \to \Rb$ such that
\begin{alignat}{2}
E(v) &= \inf_{\rho \in X} \left( F(\rho) + (v \vert \rho)\right) \quad && v \in X^\ast, \label{HKvp}\\
F(\rho) &= \sup_{v \in X^\ast} \left(E(v) - (v \vert \rho) \right) \quad && \rho \in X\label{Lvp},
\end{alignat}
where $\Rb = [-\infty,\infty]$ denotes the extended real numbers.
The function $F$ is the \emph{Lieb universal density functional}. %Convexity of the Lieb functional means that, for each pair $\rho_1,\rho_2 \in X$ and each $\lambda \in (0,1)$, we have
%\begin{equation}
%F(\lambda \rho_1 + (1- \lambda) \rho_2) \leq \lambda F(\rho_1) + (1-\lambda) F(\rho_2) \label{Fconvex}
%\end{equation}
%whereas lower semi-continuity is a weak form of continuity. Roughly speaking, lower semi-continuity of $F$ at $\rho_0$ means that $F(\rho)$ may jump up but not down as we move away from $\rho_0$.
%Although only lower semi-continuity and not continuity is guaranteed by convex conjugation, continuity is in general not precluded. It can be shown, however, that $F$ is everywhere discontinuous and therefore not differentiable.
The Lieb functional $F$ and the ground-state energy $E$ are said to be \emph{conjugate functions} or \emph{Fenchel conjugates}. Since $F$ can be calculated from $E$ and \emph{vice versa}, the two functions contain the same information, only encoded in different ways -- each property of one function is exactly reflected in some property of its conjugate function. \emph{Fenchel conjugation} is a generalization of the Legendre transformation -- the relationship between the ground-state energy and the universal density functional in DFT is thus similar to the relationship between the Hamiltonian and the Lagrangian in classical mechanics.

Comparing the Hohenberg--Kohn variation principles in Eq.\,\eqref{Fdm} and Eq.\,\eqref{HKvp}, we note that the variation is over all $X$ in Eq.\,\eqref{HKvp} but only over $\mathcal I_N$ in Eq.\,\eqref{Fdm}. If $F_\text{DM}$ is extended from $\mathcal I_N$ to $X$ by setting it equal to $+ \infty$ on $X \setminus \mathcal I_N$, then the extended $F_\text{DM}$ remains closed convex and satisfies the same Hohenberg--Kohn variation principle as does the Lieb functional $F$ in Eq.\,\eqref{Fdm}. However, by the
Fenchel--Moreau theorem, there exists only one closed convex function $F$ on $X$ that is conjugate to the closed concave function $E$ on $X^\ast$, meaning that the Lieb functional in Eq.\,\eqref{Lvp}  and the density-matrix constrained-search functional  in Eq.\,\eqref{Fdm} are the same functional~\cite{Lieb1983}:
\begin{equation}
F = F_\text{DM}.
\end{equation}
Furthermore, this functional is the lower bound to all admissible density functionals and the most well behaved such functional, being closed convex.

%Comparing with Hohenberg--Kohn theory as presented in Eqs.\,\eqref{HKvp0} and~\eqref{Lvp0}, we note the same underlying structure of convex duality and that $F_\text{HK}$ is the restriction of $F$ to ensemble $v$-representable densities.

%We conclude by comparing the Hohenberg--Kohn and Lieb variation principles in Eqs.\,\eqref{HKvp} and \eqref{Lvp} with the corresponding variation principles in Hohenberg--Kohn theory, . They are identical except for the restriction to $\rho \in \mathcal A_N \subsetneq \mathcal I_N$ and $v \in \mathcal V_N \subsetneq \mathcal X^\ast$ in the Hohenberg--Kohn theory. 

In short, DFT is essentially an exercise in convex analysis predicated on the concavity and continuity of the ground-state electronic energy $v \mapsto E(v)$. The concepts and tools of convex analysis are therefore well suited to DFT, from a computational as well as theoretical point of view.

%The convex nature of DFT is evident already in Hohenberg--Kohn theory as presented in Eqs.\,\eqref{HKvp0} and~\eqref{Lvp0} but was not developed at the time and the teaching of DFT is still mostly focused on the Hohenberg--Kohn theorem and Levy's constrained-search theory, avoiding the beautiful unifying framework of convex conjugation developed by Lieb nearly forty years ago~\cite{Lieb1983}.

%Since these sets are not explicitly known, Eqs.\,\eqref{HKvp0} and~\eqref{Lvp0} are useless from a practical point of view. We shall later examine the role of Hohenberg--Kohn theorem in Eqs.\,\eqref{HKvp} and \eqref{Lvp}.

%two  one-dimensional model systems -- a differentiable model in this section and a nondifferentiable model in the next section

\subsection{Convex conjugation: a one-dimensional differentiable model}

\begin{figure}
\centering
{\scalebox{0.75}{\includegraphics{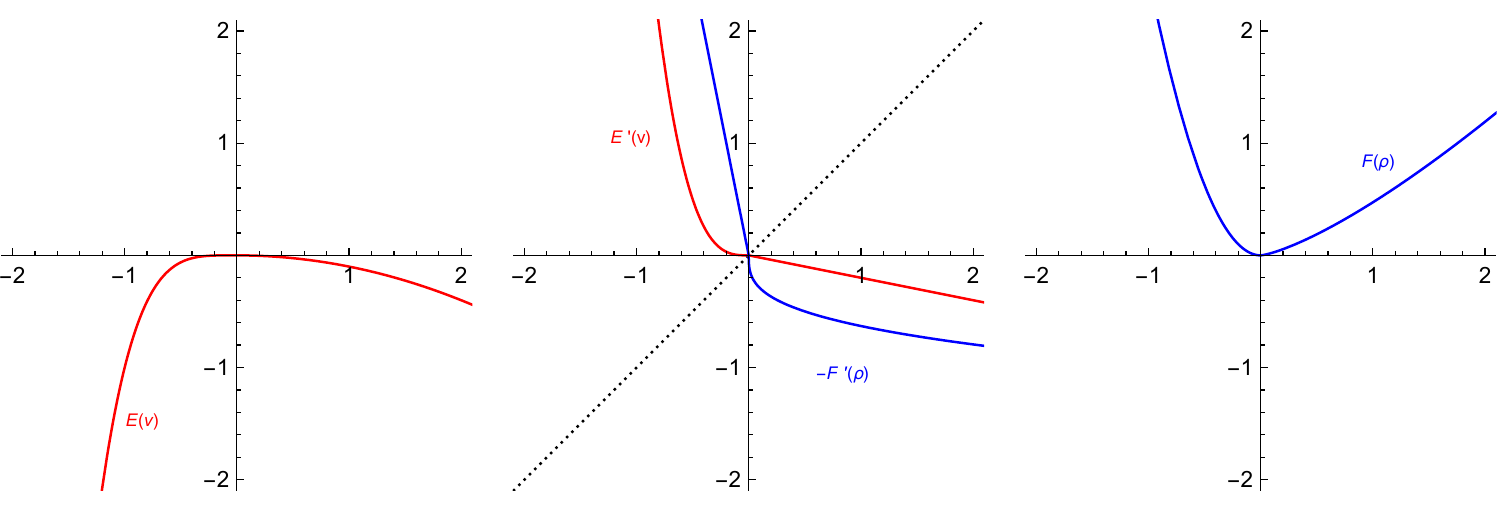}}}
\caption{A strictly concave function $\mathcal E\colon \mathbb R \to (-\infty,0]$ (left) with a strictly convex conjugate function $\mathcal F \mathbb \to [0,+\infty)$ (right). Their derivatives are plotted in the middle.}
\label{fig1}
\end{figure}

To understand  the conjugate relationship between the ground-state energy and the universal density functional better, we will consider two one-dimensional models of the energy $\mathcal E\colon \mathbb R\to \mathbb R$ and the density functional $\mathcal F\colon \mathbb R \to  \Rb$ that satisfy the \emph{Fenchel conjugations}
\begin{align}
\mathcal E(v) &= \inf_{\rho \in \mathbb R}
(\mathcal F(\rho) + ( v  \vert \rho)), \quad
\mathcal F(\rho) = \sup_{v \in \mathbb R}
(\mathcal E(v) - ( v  \vert \rho)),
\label{EFmod}
\end{align}
where the interaction between the potential and density is now the simple scalar product $(v \vert\rho) = v \rho$.
We begin  by considering in this section the differentiable functions 
plotted in Figure\,\ref{fig1}.
In agreement with the true ground-state energy, we assume that $\mathcal E \leq 0$ with $\mathcal E(0)=0$. In addition, we assume that $\mathcal E $ is twice differentiable with $\mathcal E'' < 0$ and $\lim_{v \to \pm \infty}\mathcal E'(v)= \mp \infty$. Strict concavity cannot be correct since it implies that $\mathcal E(v) < 0$ for repulsive potentials $v> 0$,  
but we accept this unphysical behaviour for the time being in order to explore the consequences of strict concavity.
%-- we shall later find that the ground-state energy is only `almost' strictly concave, which is sufficient to establish the Hohenberg--Kohn theorem.

Consider now the condition for a maximizing potential in the Lieb variation principle in Eq.\,\eqref{EFmod}.
By the derivative assumptions on $\mathcal E$, 
a maximizing potential $v \in \mathbb R$ exists for every $\rho \in \mathbb R$ as the solution to the stationarity condition $\rho = \mathcal E'(v)$ -- in this case, even for the nonphysical negative densities. Since $\mathcal E'$ is strictly decreasing, this mapping from potentials to densities can be inverted to give the Hohenberg--Kohn mapping from densities to potentials, which we recognize as the stationary condition in the Hohenberg--Kohn variation principle: $v = -\mathcal F'(\rho)$. We thus have 
two equivalent optimality conditions:
 \begin{align}
\mathcal E'(v) = \rho
\iff
- \mathcal F'(\rho) = v. \label{EFstat}
%\quad \mbox{ Hohenberg--Kohn theorem}
\end{align}
%Let us consider the stationarity conditions for the variation principles in Eq.\,\eqref{EFmod}. The stationary condition of the Lieb variation principle $\mathcal E'(v) = \rho$ can be satisfied for every $v \in \mathbb R$ by our assumptions on $\mathcal E$. Sincestationarity implies that $\mathcal E(v) = \mathcal F(\rho) + ( v \vert \rho)$, so we have two equivalent conditions:
%\begin{align}
%\mathcal E'(v) = \rho
%\iff
%- \mathcal F'(\rho) = v. \label{EFstat}
%\quad \mbox{ Hohenberg--Kohn theorem}
%\end{align}
%Since $\mathcal E$ is strictly concave with a negative second derivative, its first derivative $\mathcal E'$ exists and must be strictly decreasing. It is therefore invertible, with the inverse ${\mathcal E}^{-1} = -\mathcal F'$. 
The ground-state energy and the universal density functional are therefore \emph{functions whose derivatives (to within a negative sign) are each other's inverse functions}. These inverse relationships are illustrated in Figure\,\ref{fig1}, which shows how we may generate $\mathcal F$ by differentiation of $\mathcal E$ (left plot), inversion of $\mathcal E'$ to obtain $- \mathcal F'$ (middle plot), followed by integration of $\mathcal F^\prime$ to yield $\mathcal F$ (right plot).
Although simple, this example illustrates 
the essence of DFT.

%The external potential is thus uniquely determined by the density -- \emph{the Hohenberg--Kohn theorem thus follows from the strict concavity of the energy.} 

\subsection{Convex conjugation: a one-dimensional nondifferentiable model}

\begin{figure}
\centering
{\scalebox{0.95}{\includegraphics{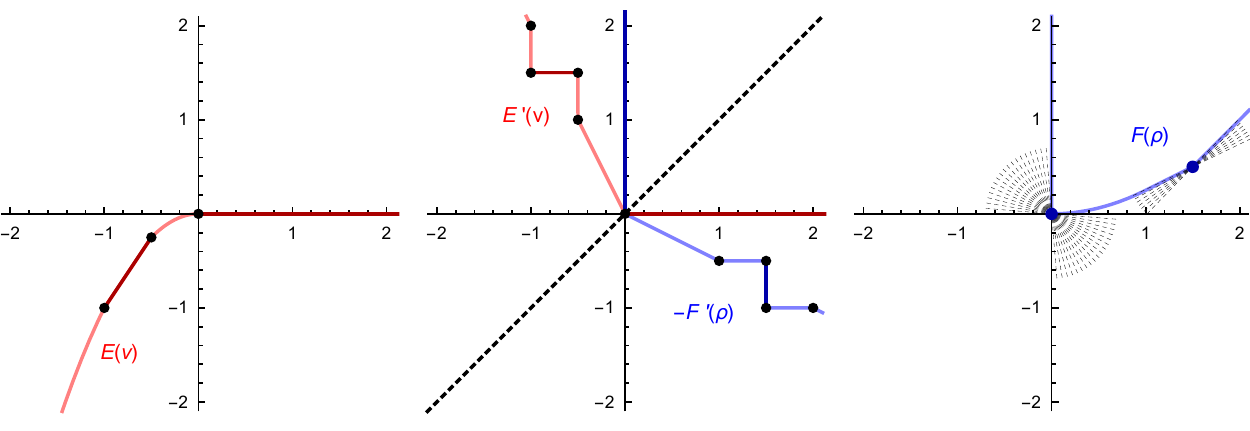}}}
\caption{A concave function $\mathcal E\colon \mathbb R \to (-\infty,0]$ (left) and its convex conjugate function $\mathcal F\colon \mathbb R \to [0,+\infty]$ (right) with their derivatives plotted in the middle.}
\label{toy2}
\end{figure}

We have seen how, in a simple model, the Hohenberg--Kohn mapping  follows from strict concavity of the ground-state energy. The true energy is not strictly concave, however, only concave.
For example, when a scalar $\mu$ is added to the potential, the energy changes linearly, in the manner $\mu \mapsto E(v+\mu) = E(v) + \mu N$ where $N$ is the number of electrons. Also, we have $E(v) = 0$ for all repulsive potentials $ v \geq 0$, thereby violating strict concavity. There may be more ways in which strict concavity is violated, but these two violations can easily be explored by a one-dimensional model.
%We also note that, unlike the simple function $\mathcal E$ considered above, the true energy $E$ is not everywhere differentiable -- in particular, it is not differentiable at potentials $v \in \mathcal V_N$ for which the ground state is degenerate. 
In Figure\,\ref{toy2}, we have plotted a  model energy $\mathcal E\colon \mathbb R \to (-\infty,0]$ that is linear on $[-1,-1/2]$ with slope 3/2 and linear on $[0,+\infty)$ with slope $0$. Also, $\mathcal E$ is nondifferentiable at $-1$ and $-1/2$. 

The effect on the density functional of this modification $\mathcal F\colon \mathbb R \to \Rb$ is dramatic. First, $\mathcal F(\rho) = + \infty$ for nonphysical densities $\rho < 0$, which is a consequence of the fact that $\mathcal E(v) =0$ for all repulsive potentials $v > 0$:
\begin{align}
\mathcal F(\rho) &= \sup_{v \in \mathbb R}
(\mathcal E(v) - ( v  \vert \rho))
\geq \sup_{v > 0}
(\mathcal E(v) - ( v  \vert \rho)) =\sup_{v > 0}
( - ( v  \vert \rho)) = + \infty.
\end{align}
Second, $\mathcal F$ becomes nondifferentiable at $\rho = 0$ and $\rho = 1/2$ -- that is, at the negative slopes of the linear segments of $\mathcal E$. From the middle plots in Figure\,\ref{toy2}, we note that each horizontal segment of $v \mapsto (v,\mathcal E'(v))$ corresponds to a vertical segment of $\rho \mapsto (\rho,\mathcal F'(\rho))$ and \emph{vice versa}, meaning that each linear segment of $v \mapsto (v,\mathcal E(v))$ corresponds to a kink of $\rho \mapsto (\rho,\mathcal F(\rho))$ and \emph{vice versa}. Roughly speaking, therefore, we may associate nondifferentiability of $\mathcal F$ with nonstrict concavity of $\mathcal E$.

Since $E$ and $F$ are nondifferentiable, we cannot express the optimality conditions in terms of derivatives, as done in Eq.\,\eqref{EFstat}. Instead, we introduce for a convex function $f\colon \mathbb R \to \Rb$ the \emph{subdifferential} of $f$ at $x$ by
\begin{equation}
\partial f(x) = [f'_-(x),f'_+(x)], \label{fsub}
\end{equation}
where $f'_-(x)$ and $f'_+(x)$ are the left and right derivatives, respectively, of $f$ at $x$. Each $m \in \partial f(x)$ is then the slope of a supporting line to $f$ and $x$, known as a \emph{subgradient} of $f$ at $x$. Returning to Figure\;\ref{toy2}, we see that $\partial \mathcal F(\rho) = \{ \mathcal F'(\rho)\}$ everywhere except at the nondifferentiable points, where
$\partial \mathcal F(0) = (-\infty,0]$ and $\partial \mathcal F(3/2) = [1/2,1]$. 
At these points, we have plotted supporting lines to $\mathcal F$ for illustration. 
%Since $0 \in \partial \mathcal F(0)$, the origin is a global minimizer of $\mathcal F$. 
For the ground-state energy, we have $\partial \mathcal E(v) = \{ \mathcal E^\prime(v) \}$ everywhere except $\partial \mathcal E(-1) = [3/2,2]$ and $\partial \mathcal E(-1/2) = [1,3/2]$.
%The maximum energy $\mathcal E(v) = 0$ is achieved for all repulsive potentials $v \geq 0$, which satisfy $0 \in \partial \mathcal E(v)$.

Subgradients allow global minima (maxima) of convex (concave) functions to be identified in a simple manner: for a convex (concave) function $f\colon \mathbb R \to \Rb$,
a point $x \in \mathbb R$ is a global minimizer (maximizer) if and only if $0 \in \partial f(x)$.
The condition for a minimum in the Hohenberg--Kohn variation principle in Eq.\,\eqref{EFmod} is therefore that the subdifferential of the convex objective function 
$\rho \mapsto \mathcal F(\rho) + (v \vert \rho)$ contains zero. Since $\partial (\mathcal F(\rho) + (v \vert \rho)) = \partial \mathcal F(\rho) + v$, this condition is equivalent to
$-v \in \partial \mathcal F(\rho)$; likewise, the condition for a global maximum in the Lieb variation principle becomes $\rho \in \partial \mathcal E(v)$. In terms of subgradients, the optimality conditions of the Hohenberg--Kohn and Lieb variation principles are therefore
\begin{equation}
 - v \in \partial \mathcal F(\rho) \iff \rho \in \partial \mathcal E(v),
\end{equation}
which reduce to the stationary conditions in Eq.\,\eqref{EFstat} when both functions are differentiable. 
%Viewing the subdifferentials as multifunctions  $\partial E\colon X^\ast \rightrightarrows X$ and $\partial F\colon X \rightrightarrows X^\ast$, 
The ground-state energy $\mathcal E$ and the density functional $\mathcal F$ are therefore \emph{functions whose subdifferential mappings $\partial \mathcal E\colon X^\ast \rightrightarrows X$ and $\partial \mathcal F\colon X \rightrightarrows X^\ast$ are each other's inverse multifunctions}, as illustrated in the middle plot of Figure\;\ref{toy2}.

%Note that, for a convex function, each local minimum is also a global minimum, greatly simplifying the optimization of convex functions.

\subsection{Hohenberg--Kohn and Lieb optimality conditions}

To set up the optimality conditions of the Hohenberg--Kohn and Lieb variation principles for the exact ground-state energy $E$ and universal density functional $F$, we must generalize subdifferentials and subgradients to functions on vector spaces. For the concave ground-state energy $E\colon X^\ast \to \mathbb R$ and the convex density functional $F\colon X \to \Rb$, the subdifferentials at $v \in X^\ast$ and $\rho \in X$ are, respectively, given by
%the density $\rho \in X$ is a subgradient at $v \in X^\ast$ if
\begin{align}
\partial E(v) &= \{ \rho \in X \,\vert \, E(\tilde v) \leq E(v) + (\tilde v -v \vert \rho), \,\forall \tilde v \in X^\ast, \, E(v) \in \mathbb R \},  \label{Esub}\\
\partial F(\rho) &= \{ v \in X^\ast \,\vert \, F(\tilde \rho) \geq F(\rho) + (v \vert \tilde \rho - \rho), \,\forall \tilde \rho \in X, \, F(\rho) \in \mathbb R \}. \label{Fsub}
\end{align}
%\begin{equation}
%E(\tilde v) \leq E(v) + (\tilde v -v \vert \rho), \quad
%\forall \tilde v \in X^\ast.
%\label{Esub}
%\end{equation}
%Likewise, for the convex density functional $F\colon X \to \Rb$, the external potential $v \in X^\ast$ is a subgradient at $\rho \in X$ if
%\begin{equation}
%F(\tilde \rho) \geq F(\rho) + (v \vert %\tilde \rho - \rho), \quad \forall \tilde %\rho \in X.
%\end{equation}
The subdifferentials $\partial E(v) \subset X$ and $\partial F(\rho) \subset X^\ast$ are convex sets, which may or may not be empty. In Eq.\,\eqref{Esub}, we recognize the subgradient inequality from Eq.\,\eqref{ineq1}, where it was defined on $\mathcal V_N \subset X$. Also, it is a simple exercise to check that $\partial F(\rho)$ in Eq.\,\eqref{Fsub} reduces to that of Eq.\,\eqref{fsub} for convex functions on the real axis.

%For an illustration, see Figure\,\ref{toy2}, where we have plotted supporting lines to $\mathcal F$ at $\rho = 0$ and $\rho = 3/2$.
% and $f'_+(x)$ are the left and right derivatives, respectively, of $f$ and $x$.   A function $f$ is differentiable at $x$ with derivative $m$ if $f$ is continuous at $x$ and $\partial f(x) = \{m \}$.
Proceeding as in the one-dimensional case, we find that the optimality conditions of the Hohenberg--Kohn and Lieb variation principles are given by
\begin{equation}
E(v) = F(\rho) + ( v \vert \rho) \iff - v \in \partial F(\rho) \iff \rho \in \partial E(v). \label{HKLopt}
\end{equation}
The subdifferential mappings $\partial F(\rho)\colon X \rightrightarrows X^\ast$ and $\partial E(v)\colon X^\ast \rightrightarrows X$ clearly play an important role in DFT, setting up the mappings between densities and potentials. 

It is instructive to consider in detail the subgradient condition on the density functional. Using the constrained-search expression for $F_\text{DM} = F$ given in Eq.\,\eqref{Fdm}, we obtain
\begin{align}
- v \in \partial F(\rho) & \iff 
 (\forall \tilde \rho \in X)\colon F(\tilde \rho) \geq F(\rho) - (v \vert \tilde \rho - \rho) \nonumber \\
 & \iff 
 (\forall \tilde \rho \in X)\colon F(\tilde \rho) + (v \vert \tilde \rho) \geq F(\rho) + (v \vert \rho) \nonumber \\
  & \iff 
 (\forall \tilde \rho \in X)\colon 
\min\nolimits_{\gamma \mapsto  \tilde \rho} \tr \left( \gamma H(v) \right)  \geq \tr \min\nolimits_{\gamma \mapsto \rho} \left( \gamma H(v) \right) \nonumber \\
&\iff 
\inf\nolimits_{\gamma} \tr \left( \gamma H(v) \right)  \geq \tr \left( \gamma_\rho H(v) \right),
\label{sub1}
\end{align}
where $\gamma_\rho$ is a ground-state density matrix of $H(v)$, thereby verifying that the conditions $-v \in \partial F(\rho) \iff \rho \in \partial E(v)$ hold if and only if $\rho \in \mathcal B_N$ is an ensemble ground-state density of $v \in \mathcal V_N$. In particular, $\rho \notin \mathcal B_N$ if and only if  $\partial F(\rho) = \emptyset$. 
By a general result of convex analysis, $F$ is subdifferentiable on a dense subset of $\mathcal I_N$. The set of ensemble $v$-representable densities is therefore dense in the set of $N$-representable densities.

Next, we  verify that $\partial F(\rho)$ is unique up to a scalar. Let $\rho$ be a ground-state density associated with the external potential $v_\rho \in \mathcal V_N$ so that $- v_\rho \in \partial F(\rho)$.
Using Eq.\,\eqref{HKLopt} and the expression for the subgradient inequality given in Eq.\,\eqref{ineq1}, we obtain
%It is interesting to see the Hohenberg--Kohn theorem arises from the subgradient inequality in Eq.\,\eqref{ineq1}. Let $\rho \in X$ and $-v_\rho \in \partial F(\rho)$. Then
\begin{align}
- v_\rho \in \partial F(\rho)& \iff \rho \in \partial E(v_\rho) \nonumber \\
&\iff \begin{cases} E(u) < E(v_\rho) + (u - v_\rho \vert \rho),  & \forall u \notin v_\rho + \mathbb R ,\\ E(u) = E(v_\rho) + (u - v_\rho \vert \rho),& \forall u \in v_\rho + \mathbb R, \end{cases}
\nonumber \\
&\iff 
\begin{cases} 
E(v_\rho) > E(u) + (v_\rho - u \vert \rho), & \forall u \notin v_\rho + \mathbb R, \\ E(v_\rho) = E(u) + (v_\rho - u \vert \rho), & \forall u \in v_\rho + \mathbb R .
\end{cases} 
\nonumber 
\end{align}
After this rearrangement of the subgradient inequality, we use Eq.\,\eqref{Esub} to give
\begin{align}
- v_\rho \in \partial F(\rho)
&\iff \begin{cases} \rho \notin \partial E(u), & \forall u \notin v_\rho + \mathbb R, \\
 \rho \in \partial E(u), & \forall u \in v_\rho + \mathbb R, \end{cases}
 \nonumber \\
 &\iff \begin{cases} -u \notin \partial F(\rho), & \forall u \notin v_\rho + \mathbb R ,\\
 -u \in \partial F(\rho), & \forall u \in v_\rho + \mathbb R, \end{cases}
 \label{sub2}
\end{align}
in accordance with the Hohenberg--Kohn theorem. Our results from Eqs.~\eqref{sub1} and~\eqref{sub2} are summarized as follows:
\begin{align}
\partial F(\rho) &= \begin{cases}
\emptyset, &  \rho \notin \mathcal B_N, \\
- v_\rho + \mathbb R , & \rho \in \mathcal B_N, 
\end{cases} \label{HKmap}\\
\!\!\partial E(v) &= \begin{cases}
\emptyset, & v \notin \mathcal V_N ,\\
\left\{ \sum_{i=1}^M\lambda_i \rho_{vi} \,\left\vert \,\sum_{i=1}^M \lambda_i = 1, \lambda_i \geq 0\right.\right\}, & v \in \mathcal V_N,
\end{cases}
\end{align}
where $v_\rho$ is an external potential such that $H(v_\rho)$ supports a ground state with ground-state density $\rho \in \mathcal B_N$, whereas $\{ \rho_{vi} \}_{i=1}^M$ are the $M$ degenerate pure ground-state densities of $v \in \mathcal V_N$. Note that $\partial E(v)$ is empty when $v$ does not support a ground state. 

\subsection{Moreau--Yosida regularization}

Although nondifferentiability is an unavoidable feature of the  density functional conjugate to the exact ground-state energy, we may instead work with the modified energy functional
\begin{equation}
E_\gamma(v) = E(v) - \frac{1}{2} \gamma \Vert v \Vert^2
\label{Emy}
\end{equation}
with $\gamma > 0$. Modified in this manner, the ground-state energy becomes strictly convex with a differentiable conjugate density functional
\begin{equation}
F_\gamma(\rho) = \min_{\tilde \rho} 
\Bigl(F(\tilde \rho) + \frac{1}{2\gamma} \Vert \tilde \rho - \rho \Vert^2 \Bigr),
\end{equation}
which is the \emph{Moreau--Yosida regularization} of the Lieb functional~\cite{Kvaal2014}. The \emph{smoothing parameter} $\gamma > 0$ can be adjusted, noting that $F_\gamma(\rho)$ approaches $F(\rho)$ pointwise from below as $\gamma \to 0$ from above. It is an attractive regularization in that, for each $v$, the true energy $E(v)$ can be recovered exactly from $E_\gamma(v)$ according to Eq.\,\eqref{Emy}. 

A slight complication of the Moreau--Yosida regularizaton is that Coulomb potentials are not square-integrable, as required to construct $E_\gamma$. This problem may be avoided by putting the system in a large box or by truncating the Coulomb potential at large separations.
In any case, Moreau--Yosida regularization shows that, even though the exact density functional is not differentiable, it is `almost' differentiable, justifying the construction of differentiable density-functional approximation.

\subsection{Four-way correspondence of density-functional theory }

An important aspect of Lieb's approach to DFT is its flexibility in describing extended density functional theories. In particular, we consider two cases, orbital-free DFT and DFT in the presence of magnetic fields. In both of these cases, the energy becomes a bifunctional with dependence on additional variables. In \emph{orbital-free DFT}, the density functional in the context of grand canonical ensemble DFT depends on both the charge density $\rho$ and the chemical potential $\mu$. For DFT in the presence of an external magnetic field, two approaches have been suggested: \emph{magnetic-field density-functional theory (BDFT)}~\cite{Grayce1994,Salsbury1997}, in which the density functional depends on the charge density $\rho$ and the external vector potential associated with the magnetic field $\mathbf{A}$; and \emph{current-density functional theory (CDFT)}~\cite{Vignale1987,Vignale1988a}, in which the density functional depends on the charge density $\rho$ and the paramagnetic component of the current density $\mathbf{j}_\text{p}$. A useful concept for these extended theories is the \emph{four-way correspondence of DFT}, which helps to identify the relationships between the bifunctionals in each of these approaches. 

\begin{figure}
\includegraphics[width=\textwidth]{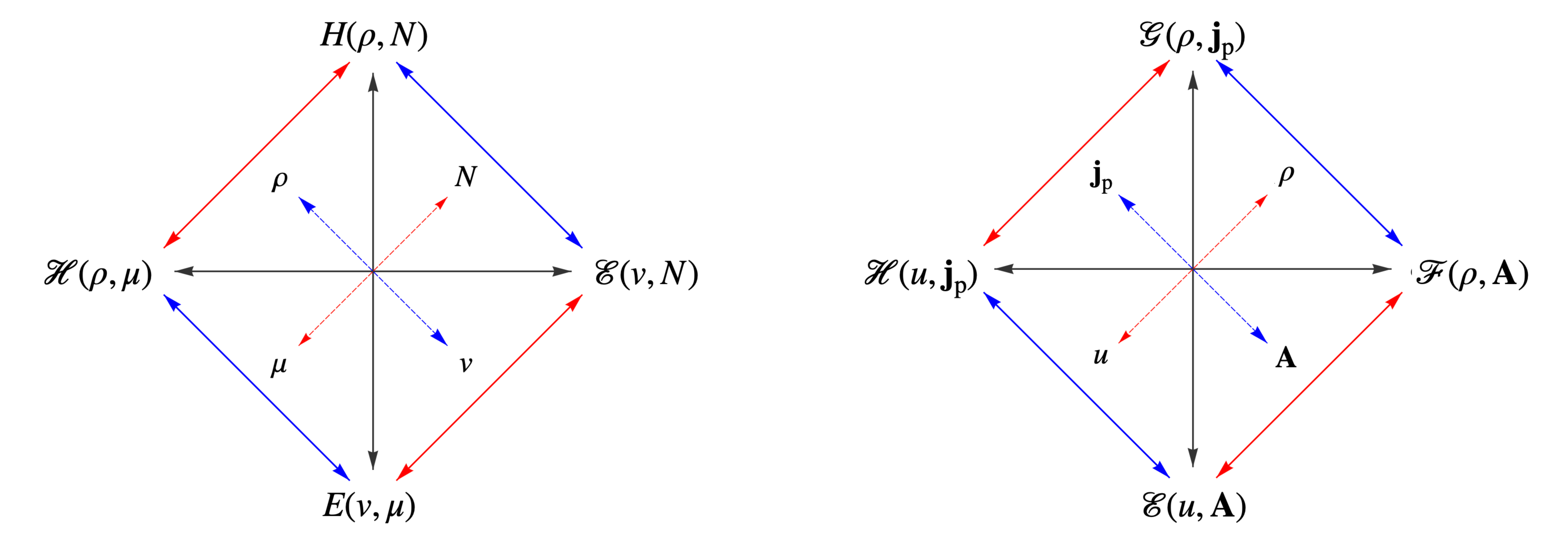}
\caption{The four-way correspondence for orbital-free DFT (left) and DFT in a magnetic field (right). The black horizontal and vertical arrows represent the relationships between each functional by bi-conjugation of both variables simultaneously. The solid diagonal arrows indicate skew conjugations of one of the variables independently. The dashed blue and red arrows indicate the dual relationships between the relevant variables.}\label{fig:fway}
\end{figure}

Figure~\ref{fig:fway} shows the four-way correspondence for orbital-free DFT in the left panel and for DFT in a magnetic field in the right panel. 
In orbital-free DFT, the energy $\mathcal{E}(v, N)$ is a functional of the external potential $v$ and the particle number $N$. The corresponding density functional depends on the charge density $\rho$ and the chemical potential $\mu$. This highlights the nature of the optimization problem in orbital-free DFT as a convex--concave saddle function. In Ref.~\citenum{Ryley2021}, an optimization scheme was developed explicitly accounting for the saddle nature of the optimization, resulting in more rapid convergence for finite systems with all electrons included.  

For DFT in the presence of magnetic fields, the four-way correspondence in Figure~\ref{fig:fway} clarifies the connections between the functionals involved in BDFT and CDFT. In particular, the variation principle used in BDFT corresponds to the partial conjugation $\rho \rightarrow u$ from $\mathcal{F}$ to $\mathcal{E}$, whilst in CDFT the variation principle corresponds to the full conjugation $(\rho, \mathbf{j}_\text{p}) \rightarrow (u, \mathbf{A})$ from $\mathcal{G}$ to $\mathcal{E}$ -- see Ref.~\citenum{Reimann2017} for a more detailed discussion of these functionals. The full mathematical characterization of CDFT using convex analysis similar to Lieb's original formulation for standard DFT has recently been completed~\cite{Tellgren2012,Kvaal2021}.

In both of these cases, the extended density functional theories can be described using the techniques of convex analysis as in Lieb's pioneering 1983 paper~\cite{Lieb1983}. In doing so, key features of the mathematical structure of the approaches and their inter-relations are clearly revealed.

\section{Applications of the Lieb variation principle}
\label{secB}

The work of Lieb to establish the variation principle of Eq.\,(\ref{Lvp}) has led not only to a clear and rigorous framework for understanding the concepts underpinning DFT, but also to a wide range of practical applications. In this section, we review various applications of the Lieb variation principle, demonstrating how it may be utilized to calculate the Kohn--Sham potential of atoms and molecules, to study the adiabatic connection and calculate the exchange--correlation functional from high-precision many-body methods, and to study the exchange--correlation hole and energy densities of atoms and molecules. 

We here pragmatically treat the density functional as differentiable, both as a function of the density and of the interaction strength. We justify this practice by noting that all approximate density functionals are taken to be differentiable and that the exact density functional is `almost differentiable' in the sense that it may be approximated to any accuracy by a differentiable Moreau--Yosida regularized functional~\cite{Kvaal2014}.

\subsection{Studying the Kohn-Sham noninteracting system}
Consider a simple generalization of the the Hamiltonian in Eq~(\ref{eq:Ham1}) to include an interaction-strength dependent two-electron interaction,
\begin{align}
  H_\lambda(v) = T + W_\lambda + \sum_{i=1}^N v(\mathbf r_i) \label{eq:HamLam}
\end{align}
where the \emph{interaction strength} between two electrons $i$ and $j$ is modulated by a parameter $\lambda \in \mathbb R$,
\begin{align}
W_\lambda = \sum_{i<j} w_\lambda(r_{ij}), \hspace{0.2in} w_0(r_{ij}) = 0, \hspace{0.1in} w_1(r_{ij}) = 1/r_{ij}, \label{eq:TwoElam}
\end{align}
here chosen such that $\lambda=0$ corresponds to a noninteracting system of electrons and $\lambda=1$ to the Hamiltonian for the usual physical system. The ground-state energy associated with this Hamiltonian is $E_\lambda$ and the corresponding interaction-strength dependent Lieb variation principle~\cite{Lieb1983} is 
\begin{align}
F_\lambda(\rho) &= \sup_{v \in X^\ast} \left(E_\lambda(v) - (v \vert \rho) \right) \label{eq:LiebVPlam}
\end{align}
This interaction-strength dependent functional provides a powerful tool to study electronic systems at any interaction strength. In particular, in Kohn--Sham theory, we are often concerned with a noninteracting system of electrons ($\lambda=0$) having the same density as that of the physical system ($\lambda = 1$). The properties of this system can be readily studied using Eq.\,(\ref{eq:LiebVPlam}) simply by setting $\lambda=0$ and using an accurate density for the physical system as input to $F_0(\rho)$. 

Savin and Colonna~\cite{Colonna1999,Savin2003} presented the first practical calculations of the Lieb variation principle in finite basis sets for atomic systems. In their approach, the optimization of the external potential $v$ was carried out directly using the derivative-free Nelder--Mead method. To facilitate this optimization, the external potential $v_\text{ext}(\mathbf r)$ was expressed in terms of a basis-set expansion with an additional $C/r$ term, whose constant $C$ was determined to account for the asymptotic form of the potential with increasing distance $r$ from the nucleus. Following this work, Wu and Yang~\cite{Wu2003} employed a similar expansion for the external potential,
\begin{align}
    v(\mathbf{r}) = v_\text{ext}(\mathbf{r}) + v_\text{ref}(\mathbf{r}) + \sum_t b_t g_t(\mathbf{r}) \label{eq:vexp}
\end{align}
where $v_\text{ext}(\mathbf{r})$ is the external potential due to the nuclei as usually evaluated, $v_\text{ref}(\mathbf{r})$ is a choice of reference potential selected to capture the long-range asymptotic decay of the potential, and the final term is a linear combination of Gaussian basis functions $g_t(\mathbf{r})$ with coefficients $b_t$ that are to be determined using the Lieb variation principle of Eq.\,\eqref{eq:LiebVPlam}. This expansion is convenient for implementation in finite-basis molecular codes and allows for direct optimization with respect to $b_t$ using derivative-based (quasi-)Newton approaches. Calculations at the $\lambda=0$ limit were presented in Ref.~\citenum{Wu2003}, with significantly accelerated convergence relative to the Nelder--Mead approach of Ref.~\citenum{Colonna1999}. Many choices of reference potential have been considered for use in Eq.\,\eqref{eq:vexp}. Commonly, the Fermi--Amaldi potential~\cite{FA} has been used. However, since this potential is not size-consistent, alternatives such as the Slater~\cite{Sla51}, Krieger--Li--Iafrate~\cite{KLI}, and localized-Hartree--Fock~\cite{LHF} exchange potentials have also been utilized as size-consistent alternatives.

The noninteracting case of Eq.\,\eqref{eq:LiebVPlam} is somewhat computationally simpler than its (partially) interacting counterparts. To see this, consider the $\lambda=0$ limit of $H_\lambda(v)$ in Eq.\,\eqref{eq:HamLam}; in this limit, the electronic Schr\"odinger equation becomes separable into a set of one-electron equations
\begin{align}
    \left[ -\tfrac{1}{2}\nabla_i^2 + v(\mathbf{r}) \right] \varphi_i(\mathbf{r}) = \varepsilon_i \varphi_i(\mathbf{r}). \label{eq:NISE}
\end{align}
which may be solved by diagonalization to give the canonical molecular orbitals $\varphi_i(\mathbf{r})$ and orbital energies $\varepsilon_i = \left \langle \varphi_i | -\frac{1}{2}\nabla_i^2 + v(\mathbf{r}) | \varphi_i \right \rangle $. The total noninteracting energy is then simply $E_0(v) = \sum_i^{n_\text{occ}} \varepsilon_i$ with the associated noninteracting wave function $\Phi_0 = \det{\left| \varphi_1 \dots \varphi_{n_\text{occ.}}\right|}$. Substitution of this expression into Eq.\,\eqref{eq:LiebVPlam} for a $v$-representable density $\rho$
leads to
\begin{align}
    F_0(\rho) % &= \max_{v \in X^\ast}(E_0 - (v|\rho)) \nonumber \\
    & = \max_{v \in X^\ast} \left( \sum\nolimits_{i=1}^{n_\text{occ}} \left \langle \varphi_i \left | -\tfrac{1}{2}\nabla_i^2 \right | \varphi_i \right \rangle  - (v|\rho_{\Phi_0} - \rho) \right) \label{eq:NILVP},
\end{align}
where $\rho_{\Phi_0}$ is the density associated with $\Phi_0$.
The computational advantage of this form for $F_0(\rho)$ is clear since no two-electron integrals are required and no choice of (partially) interacting wave-function ansatz is required. Instead, all operations amount to evaluating matrix elements of one-electron operators and solving Eq.\,\eqref{eq:NISE} with the potential of Eq.\,\eqref{eq:vexp} at each step of the optimization procedure used to evaluate Eq.\,\eqref{eq:NILVP}. At convergence, $\rho_{\Phi_0} = \rho$ and the Lieb functional becomes the noninteracting Kohn--Sham kinetic energy, $F_0(\rho) = T_\text{s}(\rho)$. Also, for $\lambda=0$, the derivative ${\partial^2 F_0(\rho)}/{\partial b_t \partial b_u }$ takes a simple form and so second-order optimization schemes such as the trust-region Newton method can be readily employed, further accelerating convergence.

Given a sufficiently accurate input density $\rho$, which may be obtained from wave-function-based quantum-chemical methods, Eq.\,\eqref{eq:NILVP} yields accurate estimates of the Kohn--Sham noninteracting kinetic energy $T_\text{s}(\rho)$, the Kohn--Sham orbitals $\varphi_i$ and eigenvalues $\varepsilon_i$, and the Kohn--Sham effective potential $v_\text{s}(\mathbf{r})$, which may be identified with the potential of Eq.\,\eqref{eq:vexp} at convergence. 
However, two complications arise for finite (orbital) basis-set calculations.

Firstly, as pointed out by Harriman~\cite{Harriman1,Harriman2}, in a finite orbital basis, there is no Hohenberg--Kohn theorem. As a result, the optimizing potential in Eq.\,\eqref{eq:LiebVPlam} may not be unique and strict convexity of $F_\lambda(\rho)$ cannot be assumed. In a complete orbital basis, the potential is determined up to a constant, leading to a well-defined shape of the Kohn--Sham potential from Eq.\,\eqref{eq:NILVP}, for example, and the constant can then be chosen arbitrarily, usually so that the potentials vanish asymptotically. In finite basis sets, the extra nonuniqueness of the potential manifests itself in a near singular behaviour of ${\partial^2 F_0(\rho)}/{\partial b_t \partial b_u }$ and convergence to oscillatory potentials. 

Secondly, in finite basis sets, an input density arising from a correlated wave function may not be exactly represented by the density of a single Slater determinant in the same basis set; see Refs.~\citenum{Harriman1,Harriman2}. In practical calculations, this may manifest itself as $|\rho_{\Phi_0}(\mathbf{r}) - \rho(\mathbf{r})| > 0$ for some points $\mathbf{r}$. It is therefore necessary to carefully check the results of calculations to ensure that any residual density differences are sufficiently small. In practice, we find that, for small atomic and molecular systems, Gaussian basis sets of augmented triple-zeta quality are adequate to obtain close approximations to correlated input densities and that this problem is essentially removed for basis sets of quadruple-zeta quality and beyond.

To avoid oscillatory potentials in calculations in finite basis sets, it is necessary to perform calculations in a regularized manner. Strategies for regularization include using approaches such as singular-value decomposition or Tikhonov regularization in second-order optimizations to avoid issues with the singular components of ${\partial^2 F_0(\rho)}/{\partial b_t \partial b_u }$. An alternative approach, which emphasizes smoothness of the calculated potentials is the smoothing-norm approach of Ref.~\citenum{HeatonBurgess2007}. In this approach, a quadratic penalty function is applied to the energy that penalizes oscillatory solutions. This approach has the advantage that the objective function, gradient, and Hessian are then consistently modified leading to simple implementation with essentially any optimization procedure. Interestingly, this approach is closely related to the Moreau--Yosida regularization of convex analysis; see the discussion in Ref.~\citenum{Kvaal2014}. Finally, we note that a similar regularization can be achieved by tailoring the basis set chosen for use in the expansion of Eq.\,\eqref{eq:vexp} -- see, for example, Ref.~\citenum{Hesselmann2007}. Whilst all of these approaches are effective in avoiding oscillatory solutions, the results of the calculations must be carefully assessed to ensure that the density $\rho_{\Phi_0}(\mathbf{r})$ remains sufficiently close to the input density $\rho$ and that the associated potentials are not over-regularized.  

\begin{figure}
\includegraphics[width=\textwidth]{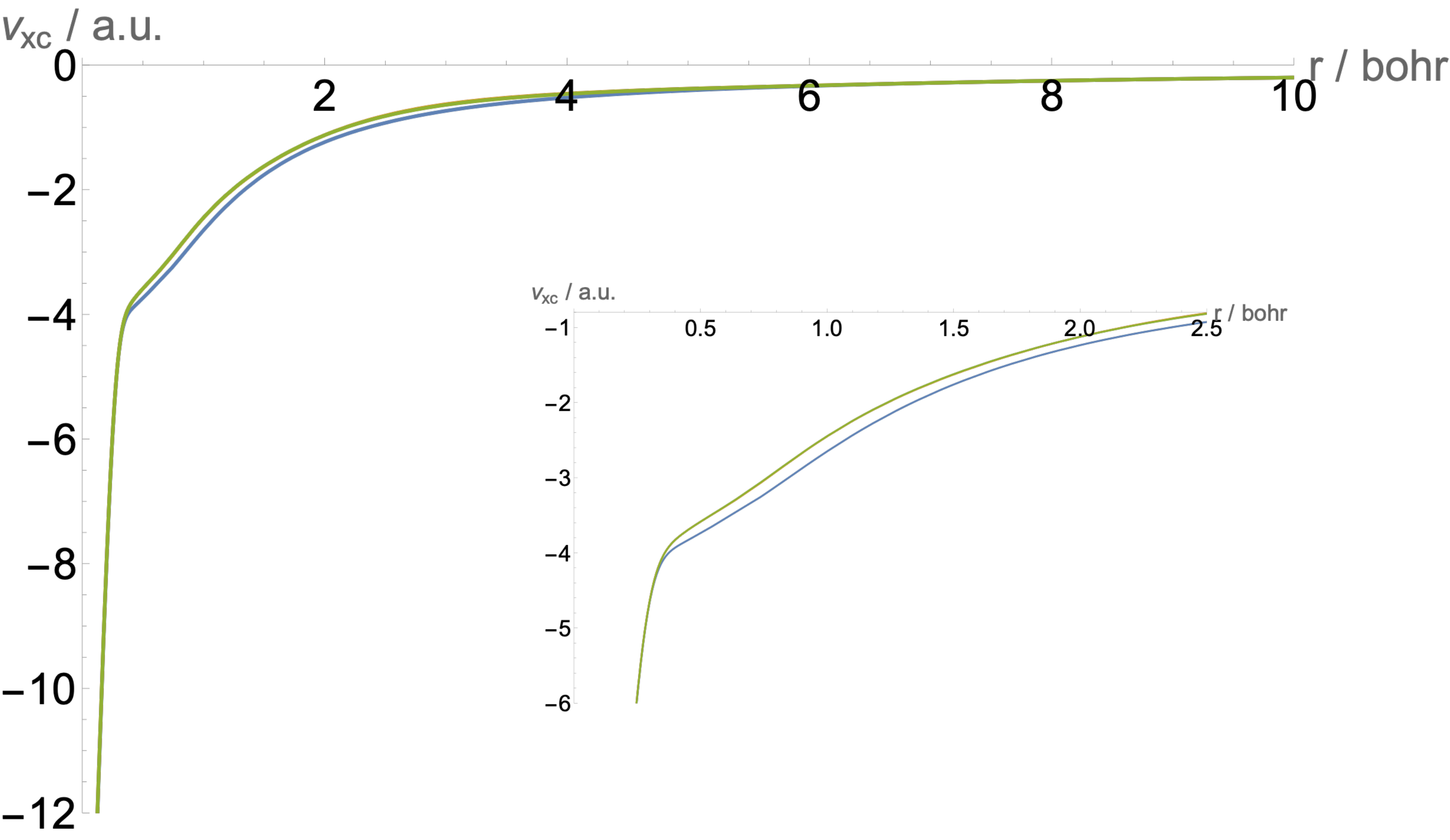}
\caption{The exchange--correlation potentials obtained by performing the Lieb optimization at $\lambda=0$ for the neon atom using the Hartree--Fock density (blue) and CCSD(T) density (green) in the uncontracted aug-cc-pCVQZ basis set. The inset highlights the differences in the valence region arising from the treatment of electron correlation. The potentials are determined using the smoothing-norm approach with a regularization parameter of $10^{-5}$ a.u.}\label{fig:xc}
\end{figure}

Figure~\ref{fig:xc} shows the exchange--correlation potentials obtained from the Lieb variation principle at $\lambda=0$. The main features of the potential are clearly captured including the $-1/r$ asymptotic decay and inter-shell structure. The inset shows the valence region, where the potentials reproducing the electron density calculated from Hartree--Fock theory and coupled-cluster single--doubles--perturbative-triples (CCSD(T)) theory are most different.  

\subsection{Lieb calculations of the adiabatic connection}\label{sec:accalc}

The adiabatic connection is a central concept in DFT, connecting the noninteracting Kohn--Sham system with the fully interacting physical system. The \emph{density-fixed adiabatic connection}, in particular, can be determined by considering how the $\lambda$-interacting Lieb functional may be expressed in terms of its noninteracting limit and a correction term,
\begin{align}
    F_\lambda(\rho) = F_0(\rho) + \int_0^\lambda \!\! F_\nu^\prime
    (\rho) \mathrm{d} \nu = T_\text{s}(\rho) + \int_0^\lambda \mathcal{W}_\nu(\rho) \mathrm{d} \nu 
\end{align}
where the prime indicates differentiation with respect to interaction strength $\nu$. Expressing $F_\lambda(\rho)$ in its constrained-search form,
\begin{align}
    F_\lambda(\rho) = \min_{\gamma \mapsto \rho} \tr \left( \gamma H_\lambda(0)\right) = \tr \left(\gamma_\lambda^\rho H_\lambda(0) \right),
\end{align}
differentiating with respect to the interaction strength, and using the Hellmann--Feynman theorem, we obtain for the linear adiabatic-connection path with $w_\lambda(r_{ij}) = \lambda/r_{ij}$ in Eq.\,\eqref{eq:TwoElam}
the \emph{adiabatic-connection integrand}
\begin{align}
    \mathcal{W}_\lambda(\rho) = \tr \left(\gamma_\lambda^\rho W_\lambda^\prime \right).
\end{align}
The Lieb variation principle therefore allows us to calculate the density-fixed adiabatic connection directly, by performing calculations for $0 \leq \lambda \leq 1$ using Eqs.~(\ref{eq:LiebVPlam}) and (\ref{eq:vexp}) as discussed in Refs.~\citenum{Colonna1999,Savin2003,Wu2003,Teale2009,Teale2010a,Teale2010,Stromsheim2011,Teale2012}.   

The integrand $\mathcal{W}_\nu(\rho)$ accounts for all electron--electron interactions. It is customary to decompose this quantity further into the \emph{classical Coulomb energy},
\begin{align}
    J_\lambda(\rho) = \frac{1}{2} \iint \!\! w_\lambda(r_{12}) \rho(\mathbf{r}_1) \rho(\mathbf{r}_2) \,\mathrm{d} \mathbf{r}_1 \,\mathrm{d} \mathbf{r}_2 ,\label{eq:EJ}
\end{align}
the \emph{exchange energy}
\begin{align}
 E_{\text{x}, \lambda}(\rho) = \tr \bigl(\gamma_0^\rho W_\lambda\bigr)  - J_\lambda(\rho),
\end{align}
and the \emph{correlation energy}
\begin{align}
E_{\text{c}, \lambda}(\rho) = \int_0^\lambda \!\!\mathcal{W}_{\text{c}, \nu}(\rho)\, \mathrm{d} \nu, \quad 
\mathcal{W}_{\text{c}, \nu}(\rho) = \tr \bigl((\gamma_\nu^\rho - \gamma_0^\rho ) W^\prime_\nu\bigr) , \label{eq:ECorAC}
\end{align}
so that 
\begin{align}
    F_\lambda(\rho) = F_0(\rho) 
    %+ \int_0^\lambda \!\! \mathcal{W}_\nu(\rho) \,\mathrm{d} \nu = 
    +J_\lambda(\rho) + E_{\text{x}, \lambda}(\rho) + E_{\text{c}, \lambda}(\rho)
\end{align}
Since the integrand is the derivative of each component with respect to the interaction strength, we see that, for the linear adiabatic connection path with $w_\lambda(r_{ij}) = \lambda/r_{ij}$, the contributions from the classical Coulomb energy and the exchange energy are constants.
The shape of the adiabatic connection is therefore entirely determined by the correlation contribution of Eq.\,\eqref{eq:ECorAC}. 

In fact, the shape of the linear-path adiabatic connection can be well understood from a perturbative analysis at $\lambda=0$ using \emph{G\"orling--Levy (GL) perturbation theory}~\cite{LevGor-PRA-94}. The correlation energy can be expanded as
\begin{align}
    E_{\text{c}, \lambda}(\rho) = \sum_{n=2}^\infty \lambda^n E_{\text{GL}}^{(n)}(\rho)
\end{align}
where $E_{\text{GL}}^{(n)}$ is the $n$th-order GL correlation energy, for which explicit expressions may be readily derived and implemented. In particular, we see that the expansion of  $\mathcal{W}_{\text{c}, \lambda}(\rho) =  \mathcal{W}_\lambda(\rho) - J_\lambda(\rho) - E_{\text{x}, \lambda}(\rho)$ is given by
\begin{align}
    \mathcal{W}_{\text{c}, \lambda}(\rho) = \sum_{n=1}^\infty \lambda^n \mathcal{W}_{\text{GL}}^{(n)}(\rho) = \sum_{n=1}^\infty \lambda^n (n+1) E_{\text{GL}}^{(n+1)}(\rho), \label{eq:GLAC}
\end{align}
from which is it clear that the slope of the linear-path adiabatic connection is $2 E_{\text{GL}}^{(2)}(\rho)$. 
%Higher-order polynomial dependence of $\mathcal{W}_\lambda(\rho)$ on $\lambda$ signifies the importance of higher-order contributions and as such is an indication of the relative significance of the correlation contributions. 
For dynamically correlated systems, where low-order perturbative expansions provide a good approximation to $E_{\text{c}, \lambda}(\rho)$, plotting $\lambda \mapsto \mathcal{W}_\lambda(\rho)$ gives smooth, almost linear curves. For systems with more significant static correlation (\emph{i.e.}, where a single Kohn--Sham determinant is not close to the physical multi-determinantal wave function), the corresponding plots display significantly more curvature.  

The Lieb variation principle is therefore a powerful tool to study not only the Kohn--Sham noninteracting system, but also the entire adiabatic connection, giving access to all of the associated $\lambda$-interacting energies, wave functions and associated density matrices. To perform practical studies of the adiabatic connection for many-electron systems in finite basis sets, a key step is to choose an appropriate, sufficiently accurate, ansatz to determine the input density $\rho$ in Eq.\,\eqref{eq:LiebVPlam} and to calculate the energy $E_\lambda$ at each interaction strength. A natural choice for few-electron systems is full configuration-interaction (FCI) theory, as employed in Refs.~\citenum{Colonna1999,Teale2009}. However, the entire repertoire of quantum-chemical methods can be utilized to study larger systems. Some care is required, however, when setting up optimization procedures to perform the Lieb maximization in Eq.\,\eqref{eq:LiebVPlam} when the ansatz for $E_\lambda$ is nonvariational, such as for M{\o}ller--Plesset and coupled-cluster theories. In particular, the required derivatives are most conveniently evaluated using the Lagrangian densities for these approaches~\cite{LagDen}; see Refs.~\citenum{Teale2009,Teale2010a} for further details.

\subsection{The adiabatic connection for H$_2$}

A simple illustrative example, for which FCI calculations can be readily performed, is the H$_2$ molecule. At short internuclear separations $R$, this system has a physical wave function that is well represented by a single Slater determinant. Correlation effects, whilst important for chemical accuracy, are relatively subtle and the CI expansion has one large coefficient on the Hartree--Fock determinant, with minor contributions from excited determinants -- characteristic of dynamic correlation. At large $R$, the system has a FCI wave function with large coefficients on more than one determinant -- characteristic of significant static correlation. At the dissociation limit, the physical wave function would be well represented by just two determinants, one for each hydrogen atom. Intermediate bond lengths capture the regime where the both dynamic and static correlation effects play a role.

\begin{figure}
\includegraphics[width=\textwidth]{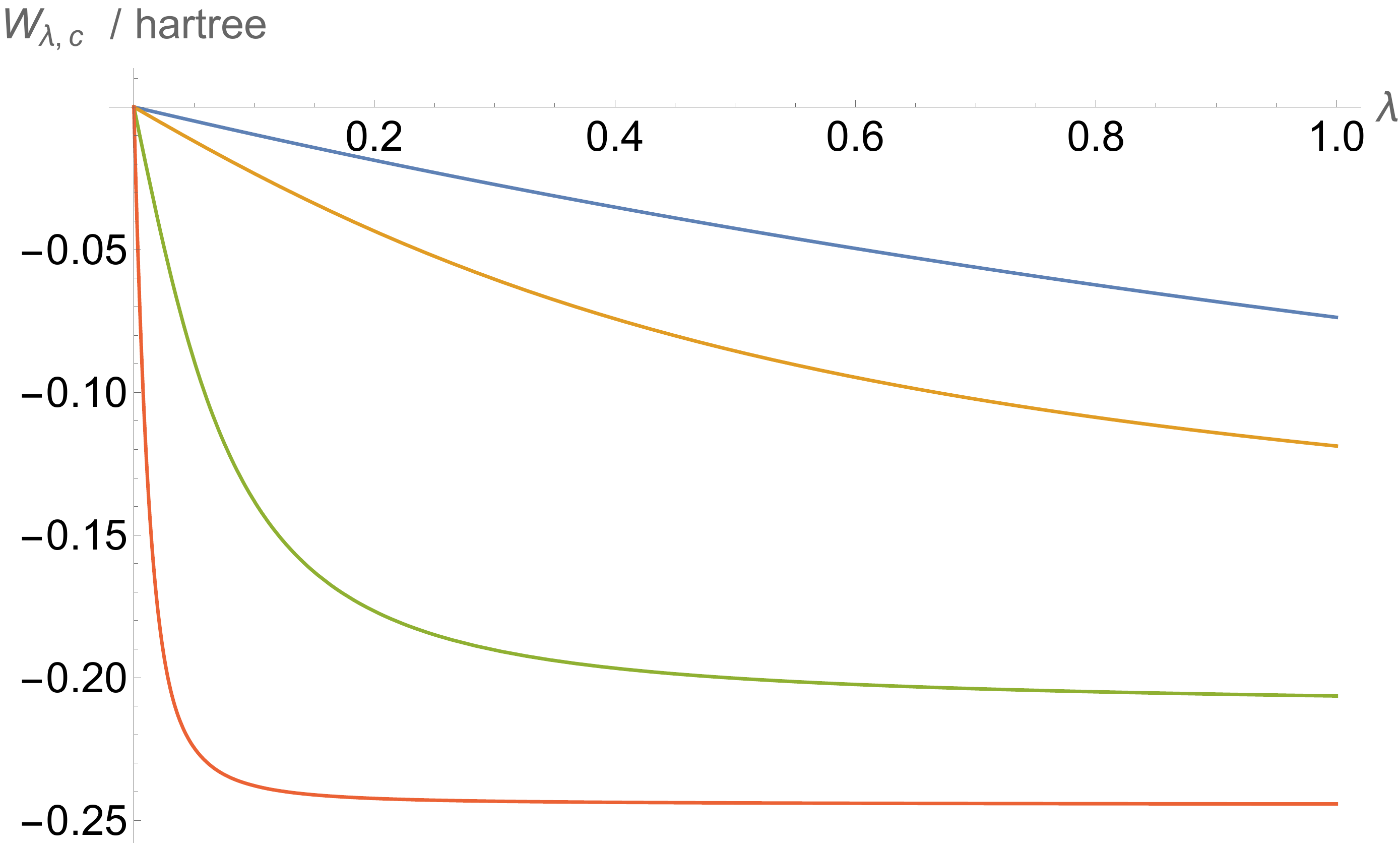}
\caption{The linear adiabatic connection for the H$_2$ molecule at internuclear separations $R = 1.4, 3.0, 5.0$ and $7.0$ bohr (top to bottom).}\label{fig:lin_ac_h2}
\end{figure}

The adiabatic-connection curves for $\lambda \mapsto \mathcal{W}_{\text{c}, \lambda}(\rho)$ calculated using the Lieb variation principle of Eq.\,\eqref{eq:LiebVPlam} at the FCI level in the uncontracted aug-cc-pCVQZ basis (for both the orbital and potential expansions) are presented in Figure~\ref{fig:lin_ac_h2} for $R = 1.4$, 3.0, 5.0 and $7.0$ bohr. These plots capture the transition from near the equilbirum geometry to almost dissociated hydrogen atoms and the corresponding transition from almost purely dynamic to static correlation. 

As expected from the perturbative analysis of Eq.\,\eqref{eq:GLAC}, the plot at equilibrium geometry displays only subtle curvature -- characteristic of low-order dynamical correlation effects. As $R$ increases, the plots exhibit more rapid curvature -- indicative of more significant higher-order correlation contributions and the relevance of static correlation. Finally, at large $R$, close to dissociation, we see a characteristic `L' curve, a feature of (almost) pure static correlation. This shape arises because the restricted Kohn--Sham single determinant cannot well represent the $\lambda$-interacting wave function, which consists of essentially one determinant for each hydrogen atom at this geometry. Since the hydrogen atoms are widely separated, there is no dynamical correlation between the electrons. As a result, the curve abruptly changes for small $\lambda$ values and then is essentially flat, reflecting the fact that, at the dissociation limit for a restricted Kohn--Sham reference wave function, $E_\text{x}(\rho) = -1/2 J(\rho)$. Therefore, to cancel the electron--electron interactions at the dissociation limit, we must have $E_\text{c}(\rho) = -1/2 J(\rho)$. The curve presented at $R = 7.0$\,bohr shows that the behaviour with respect to $\lambda$ is already approximately constant, reflecting the linear behaviour of $J(\rho)$ with respect to $\lambda$ in Eq.\,\eqref{eq:EJ}.  

\subsection{An alternative path: Generalized adiabatic connections}\label{sec:genac}

As indicated in Eq.\,\eqref{eq:TwoElam}, the noninteracting and physical systems can be connected by any number of alternative paths -- such adiabatic connections have been called generalized adiabatic connections~\cite{YangGenAC}. The Lieb variation principle of Eq.\,\eqref{eq:LiebVPlam} can for such connections be applied in the same manner as described in Section~\ref{sec:accalc} but with $E_\lambda(\rho)$ calculated using  $W_\lambda(\rho)$ defined with a different modification of the electron--electron interaction. 

\begin{figure}
\includegraphics[width=\textwidth]{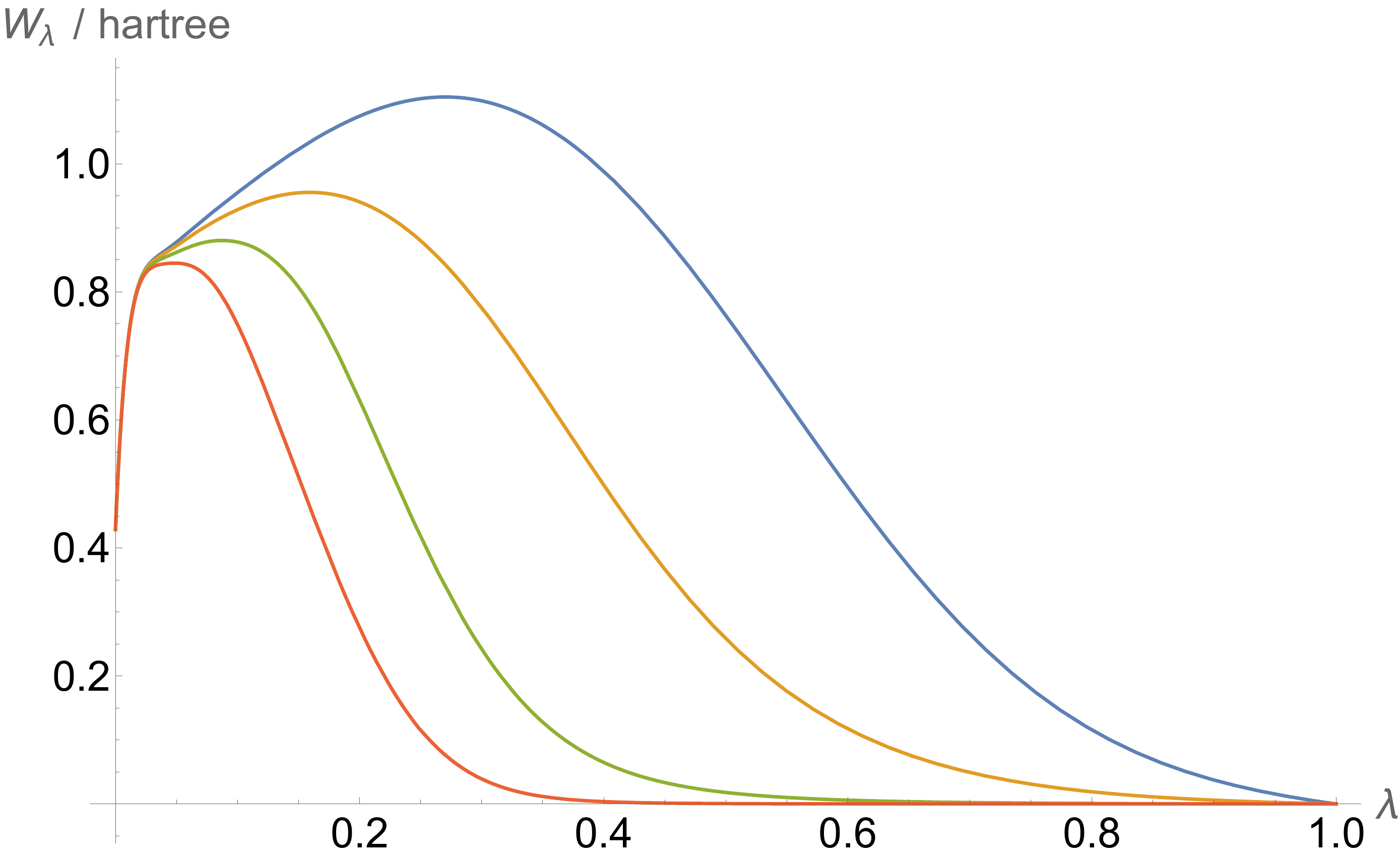}
\caption{The error-function generalized adiabatic connection for the H$_2$ molecule at internuclear separations $R = 1.4, 3.0, 5.0$ and $7.0$ bohr}\label{fig:erf_ac}
\end{figure}

In Ref.~\citenum{Teale2010}, some common adiabatic paths of relevance to range-separated methods were examined. In range-separated methods~\cite{SavinRS}, a distinction is made between long-ranged and short-ranged interactions of electrons. These interactions are typically partitioned by modifying the electron--electron interaction operator. A common approach is to use the error function to modulate the electron--electron interactions,
\begin{align}
    w_\lambda(r_{ij}) = \text{erf}\left( \mu(\lambda) r_{ij}\right) / r_{ij}, \quad \mu(\lambda) = \frac{\lambda}{1-\lambda}.
\end{align}
%Here $\mu$ is a parameter with values between $0$ and $\infty$ and a change of variables is introduced to express the interactions in terms of $\lambda$ over the range $0$ to $1$. 
%With this modulation of the electron--electron interaction, increasing $\lambda$ leads to more emphasis on short-range interactions. 
For small values of $\lambda$, the long-range interactions are captured, with the shorter-range interactions being emphasized as $\lambda$ increases. The value of the Lieb functional varies between $F_0(\rho) = T_\text{s}(\rho)$ and $F_1(\rho)$ in a manner that reflects the relative importance of different ranges of electron--electron interaction. In Figure~\ref{fig:erf_ac}, the contributions become progressively localized to small $\lambda$ values as the electron--electron interactions become predominantly long-ranged at large internuclear separation.

The Lieb variation principle provides insight into the extent to which dynamic and static correlation can be divided according to the range of interaction. By exploiting the generalized adiabatic-connection paths in conjunction with knowledge of how the generalized adiabatic connection behaves for accurate calculations, it may be possible to devise new approximations to the exchange--correlation energy that more effectively exploit range-separation techniques.  

\subsection{The exchange--correlation hole and energy densities}
\begin{figure}
\includegraphics[width=\textwidth]{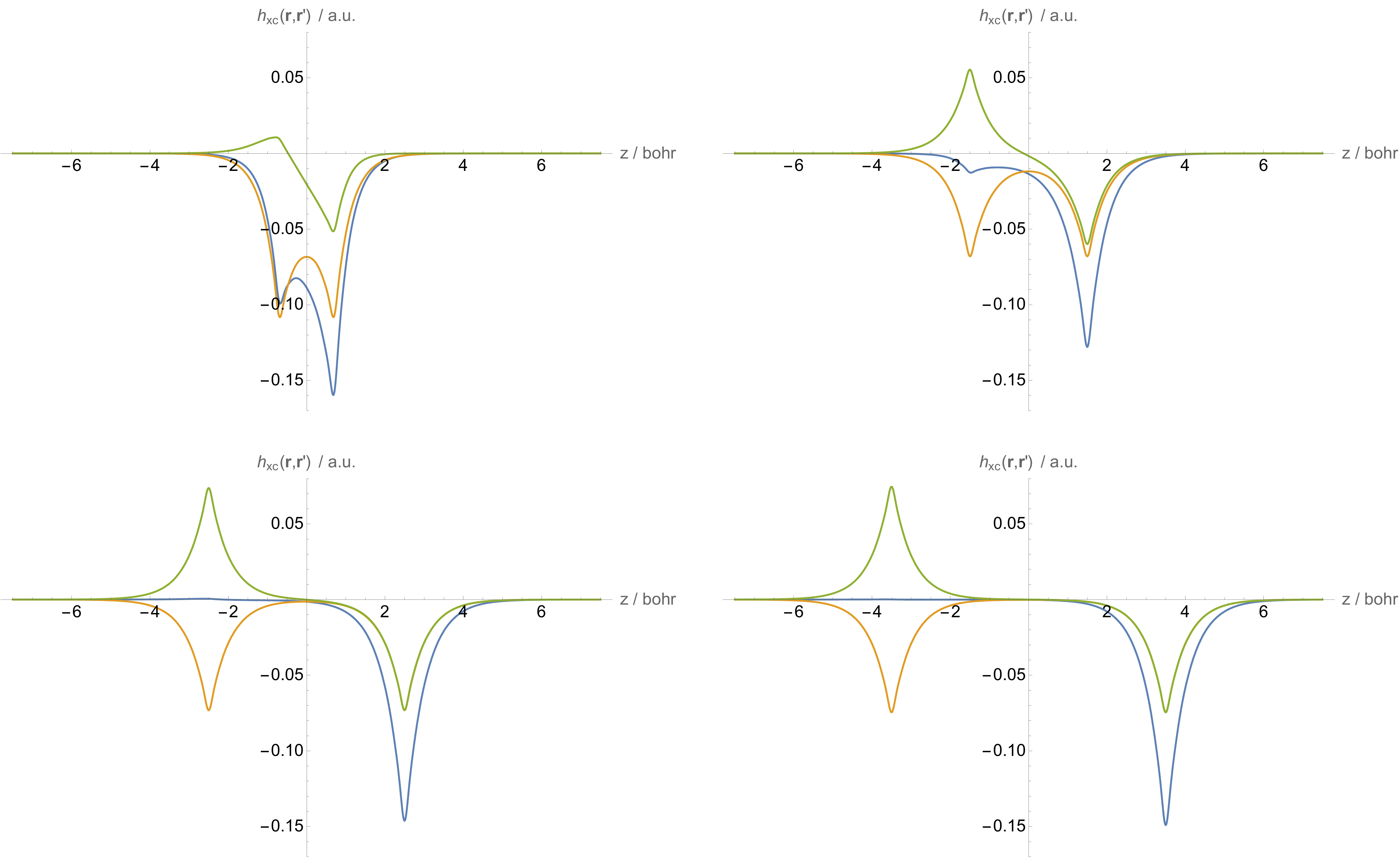}
\caption{The Fermi (orange, $\lambda=0$), Coulomb (green, $\lambda=1$) and exchange--correlation (blue, $\lambda=1$) holes for H$_2$ at internuclear separations $R = 1.4$ (top left), $3.0$ (top right), $5.0$ (lower left) and $7.0$ (lower right) bohr plotted along the internuclear axis with the position of the reference electron chosen to be 0.3 bohr left of the right-hand hydrogen nucleus.}\label{fig:xchole}
\end{figure}

The exchange--correlation hole is a central concept in DFT and its analysis has given insight into why commonly used semi-local density-functional approximations can be so successful -- see, for example, Ref.~\citenum{GritsenkoBaerends}. The exchange--correlation energy can be expressed in terms of the \emph{exchange--correlation energy density} $w_{\text{xc}, \lambda}$ as
\begin{align}
E_\text{xc}(\rho) =  \int_0^1 \!\! \int \!\! \rho(\mathbf{r}) w_{\text{xc}, \lambda}(\mathbf{r})\, \mathrm{d} \mathbf{r}\, \mathrm{d} \lambda, \quad w_{\text{xc},\lambda}(\mathbf{r}) = \frac{1}{2} \int \!\! \frac{h_{\text{xc}, \lambda}(\mathbf{r}, \mathbf{r}^\prime)}{|\mathbf{r} - \mathbf{r^\prime}|} \,\mathrm{d} \mathbf{r}^\prime.\label{eq:eden}
\end{align}   
The \emph{exchange--correlation hole} at interaction strength $\lambda$ is given by
\begin{align}
h_{\text{xc}, \lambda}(\mathbf{r}, \mathbf{r}^\prime) = \frac{P_{2, \lambda}(\mathbf{r},\mathbf{r}^\prime)}{\rho(\mathbf{r})} - \rho(\mathbf{r}^\prime)
\end{align}
where the pair density $P_{2, \lambda}(\mathbf{r},\mathbf{r}^\prime)$ is obtained using the Lieb variation principle. At $\lambda=0$, the exchange--correlation hole reduces to the Fermi (exchange) hole and so the Coulomb (correlation) hole can be identified as
\begin{equation}
h_{\text{c}, 1}(\mathbf{r}, \mathbf{r}^\prime) = h_{\text{xc}, 1}(\mathbf{r}, \mathbf{r}^\prime) - h_{\text{xc}, 0}(\mathbf{r}, \mathbf{r}^\prime).
\end{equation}
In Figure~\ref{fig:xchole}, the exchange--correlation holes for the H$_2$ molecule at internuclear separations $R = 1.4$, 3.0, 5.0 and $7.0$\,bohr are presented, plotted along the bond axis with the position of the reference electron 0.3\,bohr to the left of the second hydrogen atom, as done in Ref.~\citenum{GritsenkoBaerends}. All required quantities were calculated using the Lieb variation principle at the FCI level of theory.

At $R = 1.4$\,bohr, it is clear that the exchange hole is delocalized with a significant amplitude on both atoms. The Coulomb hole is negative close to the reference point and positive close to the other hydrogen nucleus, leading to an exchange--correlation hole that is more localized around the reference point than either of its constituent components. As the bond length increases, this effect becomes more and more pronounced, with the total exchange--correlation hole being relatively strongly localized around the reference point already at $R = 3.0$\,bohr, with only relatively modest amplitude on the left hydrogen atom. For the longer bond lengths, the localization of the exchange--correlation hole is essentially complete, with cancellation of the Fermi and Coulomb holes far from the reference point. 

This behaviour of the exchange--correlation hole rationalizes to some extent the success of popular semi-local exchange--correlation functionals. Semi-local approximations such as generalized-gradient approximations cannot be expected to capture the strong nonlocality of the Fermi and Coulomb holes. However, when exchange and correlation are treated together, the overall exchange--correlation hole is more localized and can be effectively modelled by these simple density functionals. For more detailed discussion of these ideas, see Ref.~\citenum{GritsenkoBaerends}

The exchange--correlation hole is a rather complex quantity, depending explicitly on the coordinates of two electrons. It is therefore desirable to consider quantities that reduce the complexity, whilst retaining enough information to determine the exchange--correlation energy. One possibility is to consider the energy density $w_{\text{xc}, \lambda}$ directly as a target for functional development, since one electronic coordinate has already been removed. Furthermore, it is possible to interchange the order of integration in Eq.\,\eqref{eq:eden} to generate a \emph{coupling-constant averaged energy density}, thereby also removing the $\lambda$ dependence. This simpler function can then be analysed in a similar manner to the exchange--correlation potentials generated by approximate functionals -- see, for example, Ref.~\citenum{Irons2016}. Another alternative is to consider the angle- and system-averaged exchange--correlation hole, which takes advantage of the isotropic nature of the Coulomb interaction,
\begin{align}
\bar{h}_\text{xc}(u) = \frac{1}{N} \int \!\! \rho(\mathbf{r}) \! \int \!\! h_\text{xc}(\mathbf{r}, \mathbf{r} + \mathbf{u}) \frac{\mathrm{d}\Omega_\mathbf{u}}{4 \pi} \, \mathrm{d} \mathbf{r} 
\end{align}
where $\mathbf{u} = \mathbf{r}^\prime - \mathbf{r}$ and the integration over the solid angle $\Omega_\mathbf{u}$ averages over all directions for $\mathbf{u}$. The exchange--correlation energy can then be expressed as
\begin{align}
E_\text{xc} = N \!\int_0^\infty \!\! 4 \pi u^2 \frac{\bar{h}_\text{xc}(u)}{2u} \, \mathrm{d}u
\end{align}
Both of these alternatives provide simpler functions that can be targeted in functional development. Again, the Lieb variation principle provides access to all of the quantities required to compute these functions accurately using \emph{ab-initio} methods. The resulting data can then serve as a benchmark for approximate models.

\subsection{Beyond the ground state}
The \emph{Gross--Olivera--Kohn (GOK) approach} to excitation energies~\cite{GOK1,GOK2,GOK3} in ensemble DFT can also be studied using the Lieb variation principle~\cite{LiebNATO}. In the GOK approach, a density matrix of a statistical ensemble is formed as a convex combination of the $M$ lowest states of the Hamiltonian,
\begin{align}
\gamma_{\lambda, w} = \sum_{i=1}^M w_i | \Psi_{\lambda,i} \rangle \langle \Psi_{\lambda,i}|,
\end{align}
where the ensemble weights $w_i$ satisfy $0 \leq w_{i+1} \leq w_i$ and $\sum_i w_i = 1$. The ground state is $|\Psi_{\lambda,0}\rangle$ and $|\Psi_{\lambda,i \geq 1}\rangle$ are excited states. 
%The Lieb variation principle can also be applied to this type of ensemble, see Ref.~\citenum{LiebNATO} for details. 
For an implementation of the Lieb variation principle for such ensembles, see Ref.~\citenum{Borgoo2015}. Excitation energies can be evaluated as
\begin{align}
E_1 - E_0 = (\varepsilon_1 - \varepsilon_0) + \left. \frac{\partial E_{\text{xc}, w}}{\partial w} \right|_{\rho=\rho_w}.
\end{align}  
Here $\varepsilon_1 - \varepsilon_0$ is a Kohn--Sham eigenvalue difference, which is corrected by a partial-derivative term evaluated as
\begin{align}
\left. \frac{\partial E_{\text{xc}, w}}{\partial w} \right|_{\rho=\rho^w} = \left. \frac{\partial}{\partial w} (F_{1,w}(\rho) - F_{0,w}(\rho)) \right|_{\rho=\rho_w} - J(\rho_w).
\end{align}
In this way, the Lieb variation principle may also offer insights into the behaviour of the correction term in the GOK approach to excitation energies; see Ref.~\citenum{Borgoo2015} for further details. 

Other approaches to excited states have been developed using perturbation theory along the linear and generalized adiabatic connections~\cite{Rebolini2015}, which were discussed in Sections~\ref{sec:accalc} and~\ref{sec:genac}. In these cases, the Lieb variation principle was used to determine the ground-state adiabatic connections and then excitation energies estimated by perturbation at each value of the interaction strength. The Lieb variation principle may provide a useful tool for further study of time-independent approaches to excitation energies in the future. 

\section{Conclusions}
\label{conclusions}

We have seen how DFT is predicated on the concavity and continuity of the ground-state energy in the external potential -- from this fact, DFT follows by application of convex analysis. In particular, the universal density functional and the ground-state energy are conjugate functions, containing the same information but encoded in different ways. While the ground-state energy can be obtained from the universal density functional by the Hohenberg--Kohn variation principle, the density functional can be calculated from the ground-state energy by the Lieb variation principle.

The Lieb variation principle is not only theoretically important, giving insight into the structure of DFT, but it is also a practical computational tool. It allows us to calculate, for any $N$-representable density, the universal density functional and the external potential that supports this density (if such a potential exists). The utility of this approach has been briefly demonstrated here for determining the Kohn--Sham potential, the adiabatic connection and detailed information on the exchange and correlation holes, to high accuracy using \emph{ab-initio} many-body wave function approaches. This tool can give valuable insight into the numerical behaviour of the exact universal density functional, which may serve as a useful benchmark to guide the development of approximate models.

It is our view that the beautiful convex formulation of DFT first put forward by Lieb~\cite{Lieb1983} has not yet received the attention it deserves, as a framework for teaching DFT and as a tool for further development of DFT -- we hope the present overview goes some way towards rectifying this situation. We also highlighted how Lieb's approach can be utilized for extended density-functional theories such those required for systems in a magnetic field and for orbital-free density functional theory. This further illustrates how Lieb's convex formulation of DFT can be used to unify the presentation of different variants and extensions of DFT and clarify the relationships between them. Lieb's convex formulation of DFT continues to illuminate the development of state-of-the-art approaches in DFT almost forty years after its publication.   

\begin{ack}
We thank Dr. Tom J. P. Irons and Dr. Bang Huynh for useful discussions in the preparation of this manuscript. 
\end{ack}

\begin{funding}
This work was partially supported by the Research Council of Norway through its Centres of Excellence scheme, project number 262695.
We acknowledge financial support from the European Research Council under H2020/ERC Consolidator Grant top DFT (Grant No. 772259). 
\end{funding}

%------
% Insert the bibliography.
%------

\bibliographystyle{jcp}
\bibliography{main}

%\bibliographystyle{rsc} %the RSC's .bst file

%\begin{thebibliography}{99}

%------ Example for a paper in journal:
% \bibitem{article1}
% A.~Petrunin, Parallel transportation for Alexandrov space with curvature bounded below.
% \emph{Geom. Funct. Anal.} \textbf{8} (1998), no.~1, 123--148.
% \Zbl{0903.53045} \MR{1601854}

%------ Example for a book:
% \bibitem{book1}
% W.~P. Ziemer, \emph{Weakly differentiable functions}.
% Grad. Texts in Math. 120,  Springer, New York, 1989.
%\Zbl{0692.46022} \MR{1014685}

%------ Example for a paper in a book:
% \bibitem{incollection1}
% J.~S. Milne, Introduction to Shimura varieties.
% In \emph{Harmonic analysis, the trace formula, and Shimura varieties},
% edited by M.~W. Marcellin and E.~Giorgi, pp. 265--378,
% Clay Math. Proc. 4, Amer. Math. Soc., Providence, RI, 2005.
% \Zbl{1148.14011} \MR{2192012}

%------ Example for a preprint on arXiv:
% \bibitem{preprint1}
% D.~V. Nguyen, S.~K. Chilappagari, M.~W. Marcellin, and B.~Vasic,
% LDPC codes from latin squares free of small trapping sets,
% 2010, \href{http://arxiv.org/abs/1008.4177}{arXiv:1008.4177}.

%------ Example for a report:
% \bibitem{report1}
% J.~Schöberl, Commuting quasi-interpolation operators.
% Technical report isc-01-10-math, Texas A\&M University, 2001,
% \url{www.isc.tamu.edu/publications-reports/tr/0110.pdf}.

%------ Example for a thesis:
% \bibitem{thesis1}
% E.~Giorgi, \emph{The geometric universe}.
% Ph.D. thesis, University of Maryland, College Park, 2002.

%\end{thebibliography}

\end{document}